\documentclass[aps,pra,superscriptaddress,amsmath,amssymb,preprintnumbers,floatfix,showpacs,11pt]{revtex4}

\usepackage{amssymb}
\usepackage{epsfig}

\begin{document}
\title{ On the evolution of  superposition of squeezed displaced number
states with the multiphoton Jaynes-Cummings model }
\author{ Faisal A. A. El-Orany}
 \affiliation{Department of Mathematics  and computer Science,
Faculty of Science, Suez Canal University 41522,
 Ismailia, Egypt}

 \author{and A.-S. Obada}

 \affiliation{ Department of Mathematics, Faculty of Natural Science,
Al-Azhar University, Nasr City 11884, Cairo, Egypt}

\date{\today}

\begin{abstract}
In this paper we discuss the quantum properties for
 superposition of squeezed  displaced number states against multiphoton
Jaynes-Cummings model (JCM).  In particular, we investigate atomic
inversion, photon-number distribution, purity, quadrature
squeezing, Mandel $Q$ parameter and Wigner function. We  show that
the quadrature squeezing for three-photon absorption case can
exhibit  revivals and collapses typical  to those
 occurring in  the atomic inversion for one-photon absorption case. Also we
prove that for odd number absorption parameter there is a
connection  between the evolution of  the atomic inversion and the
evolution of the  Wigner function at the origin in phase space.
Furthermore, we  show that the nonclassical states whose  the
Wigner functions values at the  origins  are negative will be
always nonclassical when they are evolving through the JCM with
even absorption parameter. Also we demonstrate that various types
of
 cat states  can be generated via  this  system.

\end{abstract}

  \maketitle

\section{Introduction}
 Jaynes-Cummings model (JCM)
\cite{jay1} is the simplest nontrivial example of two interacting
quantum systems: a two-level atom and a single mode of the radiation
field. In addition to its being exactly solvable, the physical system
that it represents has  become experimentally realizable with
the Rydberg atom in high-$Q$ microwave cavities, e.g. \cite{remp}.
Such experimental realization has stimulated an intense research  devoted
to highlighting and generalizing the original model.
For example, this model has been extended to include, e.g., multiphoton
\cite{mp1,{mp2},{maq}}, intensity-dependent coupling \cite{idc1} and
damping and dissipation \cite{damp}.
Investigation of the JCM has produced many interesting effects, which have no
classical analogue, such as
revival-collapse phenomenon in the atomic inversion \cite{eber},
quadrature squeezing \cite{meys} and sub-Poissonian statistics
\cite{kim2}. Furthermore, JCM has been used as a source for the nonclassical
states by means of the conditional measurement technique \cite{cm1}.
For more details related to JCM reader can consult \cite{bruce}.
On the other
hand, evolution of  states with JCM is an important topic, which can give
insight into the occurrence of the nonclassical effects and how can
be controlled based on the values of the interaction parameters. In this
respect we can mention, e.g., Shr\"{o}dinger-cat states
\cite{vid,{ger1},{jos}}, displaced number states \cite{kim1}
 and squeezed coherent states \cite{mil,{rice},{mann}}.

 Recently, we have introduced a general class of quantum states as a
single mode vibration of electromagnetic field suddenly
squeezed-plus-displaced by a collection of two displacements $\phi$ out
of phase with respect to each other \cite{[21],{fall}}. These states (SSNDS) are
defined as
\begin{equation}
|\alpha ,r ,n\rangle_{\epsilon} =\lambda_{\epsilon} [\hat{D}
(\alpha )+\epsilon \hat{D}(-\alpha )]\hat{S}(r)|n\rangle ,\label{1}
\end{equation}
where $\lambda_{\epsilon} $ is a normalization constant,
$\hat{D}(\alpha )$
and $\hat{S}(r)$ are displacement and squeeze operator, respectively, while
$\alpha $ and $r$ are displacement and squeeze parameters.
$\epsilon=|\epsilon|\exp(i\phi) $, where $\phi$ is the relative phase
and throughout the paper $\epsilon=0,i,1,-1$ in
correspondence with squeezed displaced number  state, Yurke-Stoler-,
 even- and odd-squeezed-displaced number states.
Also $|n\rangle $
denotes Fock state. Squeeze and displacement operators are given, respectively, by
\begin{equation}
 \hat{S}(r)=\exp[ \frac{r}{2}( \hat{a}^{2}-
\hat{a}^{\dagger2})],\qquad
 \hat{D}(\alpha )=\exp
(\hat{a}^{\dagger}\alpha-\hat{a}\alpha^{*}),
 \label{2}
 \end{equation}
 where $\hat{a}$ and $\hat{a}^{\dagger }$ are annihilation and
creation operators. The normalization constant $\lambda_{\epsilon}$ is given by

\begin{equation}
| \lambda_{\epsilon}| ^{-1}= 1+|\epsilon|^{2}+2|\epsilon| \exp
(-2|\beta|^{2} ) {\rm L}_{n}(4|\beta|^{2})\cos\phi,
\end{equation}
 with
\begin{equation}
\beta=\alpha \cosh r+\alpha^{*}\sinh r;
\label{4}
\end{equation}
 and $L_{n}(.)$ are the Laguerre polynomials.
Furthermore, the density matrix $\hat{\rho}_{f}(0)$ of this state can be written as
\begin{eqnarray}
\begin{array}{lr}
\hat{\rho}_{f}(0)=
|\psi^{(1)}_{f}(0)\rangle\langle \psi^{(1)}_{f}(0)|
\\
\\
=| \lambda_{\epsilon }|^{2}(\hat{\rho}_{M}+\hat{\rho}_{I}),  \label{501}
\end{array}
\end{eqnarray}
where  $|\psi^{(1)}_{f}(0)\rangle=|\alpha,r,n\rangle_{\epsilon}$
and the subscript $f$ stands for the radiation field density matrix.
The second line of (\ref{501}) includes the statistical mixture part
$\hat{\rho}_{M}$ and interference part $\hat{\rho}_{I}$ of the density
matrix.
The statistical mixture part has the form
\begin{equation}
\hat{\rho}_{M}=
 \hat{D}(\alpha )\hat{S}(r)|n\rangle
\langle n| \hat{S}^{\dagger}(r)\hat{D}^{\dagger}(\alpha )
+|\epsilon|^{2} \hat{D}(-\alpha )\hat{S}(r)|n\rangle
\langle n| \hat{S}^{\dagger}(r)\hat{D}^{\dagger}(-\alpha ),  \label{52}
\end{equation}
while the quantum interference part is
\begin{equation}
\hat{\rho}_{I}=
 \epsilon^{*}
 \hat{D}(\alpha )\hat{S}(r)|n\rangle
\langle n| \hat{S}^{\dagger}(r)\hat{D}^{\dagger}(-\alpha )
+\epsilon \hat{D}(-\alpha )\hat{S}(r)|n\rangle
\langle n| \hat{S}^{\dagger}(r)\hat{D}^{\dagger}(\alpha ).  \label{53}
\end{equation}
This  part, i.e. $\hat{\rho}_{I}$, of the density matrix includes information
about the quantum interference between the components of the state.
On the other hand,
in the number state basis (\ref{1}) takes the form \cite{fall,{faisal3}}

\begin{equation}
|\alpha ,r,n\rangle_{\epsilon}=\sum\limits_{m=0}^{\infty}
C_{m}(\alpha,r,n,\epsilon )|m\rangle, \label{5}
\end{equation}
where the distribution coefficient has the form
\begin{eqnarray}
\begin{array}{rl} C_{m}(\alpha, r,n,\epsilon)
=\frac{\lambda_{\epsilon} (\frac{\tanh
r}{2})^{\frac{m}{2}}}{\sqrt{n!m!\cosh r}}
 [1+(-1)^{(n+m)}\epsilon ]       \exp [\frac{\alpha ^{2}e^{2r}}{2}(\tanh r-1)]
\\
\\
\times
\sum\limits_{j=0}^{min(m,n)}\frac{n!m!}{j!(n-j)!(m-j)!}[\frac{2}{\sqrt{\sinh 2r}}]^{j}
[\frac{-\tanh r}{2}]^{\frac{(n-j)}{2}}
 {\rm H}_{n-j}\left(\frac{i\alpha}{\sqrt{\sinh (2r)}}\right)
 {\rm H}_{m-j}\left(\frac{e^{r}\alpha}{\sqrt{\sinh (2r)}}\right),
\end{array}
\label{5a}
\end{eqnarray}
whereas ${\rm H}_{m}(.)$ is the Hermite polynomial of order $m$ and
 $\alpha$ has been considered to be real. The photon-number distribution
$P(m)$  of the state (\ref{1}) is
\begin{equation}
P(m)=|C_{m}(\alpha,r,n, \epsilon)|^{2}.\label{5b}
\end{equation}

For completeness,
 it has been shown that  states (\ref{1}) can be
generated by  the so-called quantum state engineering and also by
 trapped ion \cite{[21],{fall}}.
The quantum properties of these states reveal that they can exhibit
sub-Poissonian statistics, quadrature squeezing and oscillations in
photon-number distribution.
Their phase properties  from the point of view of the Pegg-Barnett
Hermitian phase formalism \cite{faisal3}
 show that there are two regimes controlling
the behaviour based on whether
the superposition is macroscopic ($\alpha>>1$) or microscopic
($\alpha \leq 1$). For such superposition the nonclassical effects are more pronounced  in the
microscopic regime. Moreover, the influence of thermal noise on
the behaviour of (\ref{1}) has been considered  in \cite{fais}
showing that the correlation between different oscillators is essentially
responsible for the occurrence of the nonclassical effects, which is
similar to that of  Schr\"{o}dinger-cat states with thermal noise
\cite{kaic}.

For convenience we mention that a new set of squeezed states using
group-theoretical methods has been presented earlier, e.g. \cite{su1}.
Actually, the definition of this set is based on the Holstein-Primakoff
realization of both $SU(2)$ and $SU(1,1)$.  It has been shown that such type
of states can
 exhibit interesting squeezing properties, depending in a
characteristic way on the dimension of the irreducible unitary
representation adopted. Moreover, the multiphoton, many-mode
squeezed states for $SU(n)$, using a generalization Holstein-Primakoff
realization have been also reported \cite{su2}.
Finally, a long list of references related to different types of
squeezed states can be found in the recent review \cite{doda}.

In this paper we study the evolution of the states (\ref{1}) with the multiphoton
Jaynes-Cummings model.
The motivation behind this work is to determine the evolution of the
nonlinear process of the multiphoton transition in the presence of
quantum interference of two squeezed displaced number states.
Also, as we will show below, there are some intriguing
features such as  information about
atomic inversion of the  JCM can be obtained
from quadrature squeezing of the nonlinear JCM, and also
there is a connection between the evolution  of the
Wigner function at the origin in phase space and the atomic inversion.
In addition to that this work gives a generalization to several
results given in the literatures earlier, e.g. \cite{vid}--\cite{mann}.
Actually, there is some  merit in having the most general results.

The Hamiltonian controlling the system in the rotating wave
approximation (RWA) is
\begin{equation}
\frac{\hat{H}}{\hbar}=\omega_{0}\hat{a}^{\dagger}\hat{a}+
\omega_{a}\hat{\sigma}_{z}+g(\hat{a}^{k}\hat{\sigma}_{+} +
\hat{a}^{\dagger k}\hat{\sigma}_{-}),
 \label{6}
\end{equation}
where $\hat{\sigma}_{\pm}$ and $\hat{\sigma}_{z}$ are the Pauli spin
operators;
$\omega_{0}$ and $\omega_{a}$ are the frequencies of
cavity mode and the atomic transitions, respectively;  $g$ is the atom-field coupling
constant and $k$ is the absorption parameter.

Throughout the paper we assume that the atom is initially prepared  in
its  excited state $|+\rangle$ and the field is in the
SSDNS (\ref{1}). In this case the initial density matrix for the system is

\begin{equation}
\hat{\rho}(0)=|+\rangle \langle +| \bigotimes
|\psi^{(1)}_{f}(0)\rangle\langle \psi^{(1)}_{f}(0)|,  \label{7}
\end{equation}
where $\bigotimes$ stands for the direct product.
Working in the interaction picture and considering the resonance case
$\omega_{0}=k\omega_{a}$, after well-known analytical
procedures, the time evolution  density matrix describing the system is
\begin{eqnarray}
\begin{array}{lr}
\hat{\rho}(T)=|\psi (T)\rangle\bigotimes \langle \psi (T)|\\
\\
 =\sum\limits_{m,m'=0}^{\infty}
C_{m}(\alpha,r,n,\epsilon)C^{*}_{m'}(\alpha,r,n,\epsilon)
\Bigl\{\cos [T\sqrt{h(m,k)}] |m,+\rangle
 \\
 \\
- i\sin [T\sqrt{h(m,k)}] |m+k,-\rangle\Bigr\}\bigotimes
\Bigl\{\langle m',+| \cos [T\sqrt{h(m',k)}]
+i \langle m'+k,-|\sin [T\sqrt{h(m',k)}] \Bigr\},
\end{array}
 \label{8}
 \end{eqnarray}
 where $T=gt$ is the scaled time, $|-\rangle$ is atomic ground state
  and $h(m,k)=\frac{(m+k)!}{m!}$. Also $|\psi (T)\rangle$
  is the dynamical state vector of the system.
 From (\ref{8})  the entanglement of the two interacting
 quantum systems is readily apparent.
 To evaluate quantities associated with atom (or field) we have to trace
(\ref{8}) over the field (or the atom) system.
As it is obvious the system includes various parameters, which make such
analysis  difficult. For this reason throughout the paper
we restrict ourselves to the parameters, which can give
significant results.  Also throughout the text the statement
standard JCM  means $k=1$, the optical cavity field and atom are
initially prepared
 in coherent states and  in the atomic excited state, respectively.

The paper is organized as follows: In section 2 atomic inversion is
investigated.
Section 3 is devoted to the photon-number distribution and purity.
In section 4 Mandel $Q$ parameter and quadrature squeezing are discussed.
In section 5 Wigner function is demonstrated.
Conclusions are summarized in section 6.
\section{Atomic inversion}
In this section we  discuss the revival-collapse phenomenon in
the evolution of  the atomic inversion for the density matrix (\ref{8}).
This phenomenon is a pure quantum mechanical effect and having its
origin in the granular structure of the photon-number distribution of the
initial field \cite{zaber1}. Furthermore, this  phenomenon has been realized
experimentally in the sense that  the state of the atomic beam leaving the cavity
is monitored by ionization detectors \cite{remp}.
On the other hand, the revival-collapse phenomenon has been seen
also in  nonlinear optics
for the single-mode mean-photon number of the Kerr nonlinear
coupler \cite{perina1} when the modes are  initially prepared in
coherent light, however, in this case the origin of the occurrence of such
phenomenon is in the presence of nonlinearity in the system
(third-order nonlinearity specified by the cubic susceptibility).
We should
stress here that the analytical formulae of both the
 mean-photon number of the Kerr coupler
and of the atomic inversion of the JCM
(under the harmonic approximation) \cite{eber} are similar. This remark
makes the occurrence of the collapse-revival pattern in the Kerr coupler is
an expected result.

Now the atomic inversion  for the density matrix (\ref{8}) reads
\begin{eqnarray}
\begin{array}{lr}
\langle \hat{\sigma}_{z}(T)\rangle={\rm Tr}_{f}
[\hat{\sigma}_{z}(0)\hat{\rho}(T)]\\
\\
=\sum\limits_{m=0}^{\infty}P(m)\cos[2T\sqrt{h(m,k)}],\label{9}
\end{array}
\end{eqnarray}
where $P(m)$ is given by (\ref{5b}) and the subscript $f$ on the right
hand side means that we
trace over the field system.
Firstly, it is worth reminding that the behaviour of the
$\langle \hat{\sigma}_{z}(T)\rangle$ of squeezed coherent light exhibits various
interesting effects \cite{mil,{rice}}.
For instance, it has been shown that for strong  coherent contribution
 \cite{mil} the collapse time
depends on the direction of the squeezing and  for certain
values of the squeezing parameter the response of the atom is similar to that of the
chaotic radiation field.
On the contrary, for strong squeezing contribution \cite{rice}
the dynamical response of the atomic inversion shows echoes after each revival
resulting from the interference effect.
We start our discussion for $k=1$  and the cavity field
is in  superposition of displaced number states (see Figs. 1 for given values
of the parameters).
Actually, the superposition of displaced number states
can exhibit strong sub-Poissonian
statistics as well as quadrature squeezing.
Also they can be generated
via the driven Jaynes-Cummings model or via trapped ion
\cite{moya,{zeh1}}.
We proceed, it has been shown that there is  a close correspondence
between the number of peaks
(with  Poissonian envelopes) involving in $P(m)$
and the number of occurred revivals  in $\langle \hat{\sigma}_{z}(T)\rangle$
\cite{kim1}.
In other words,
 the envelope of each revival is a readout of the
photon distribution, in particular, for slowly varying photon distribution
\cite{fleisch}
 provided that the atom is in the excited or in the ground state.
\begin{figure}
{\includegraphics[width=5cm]{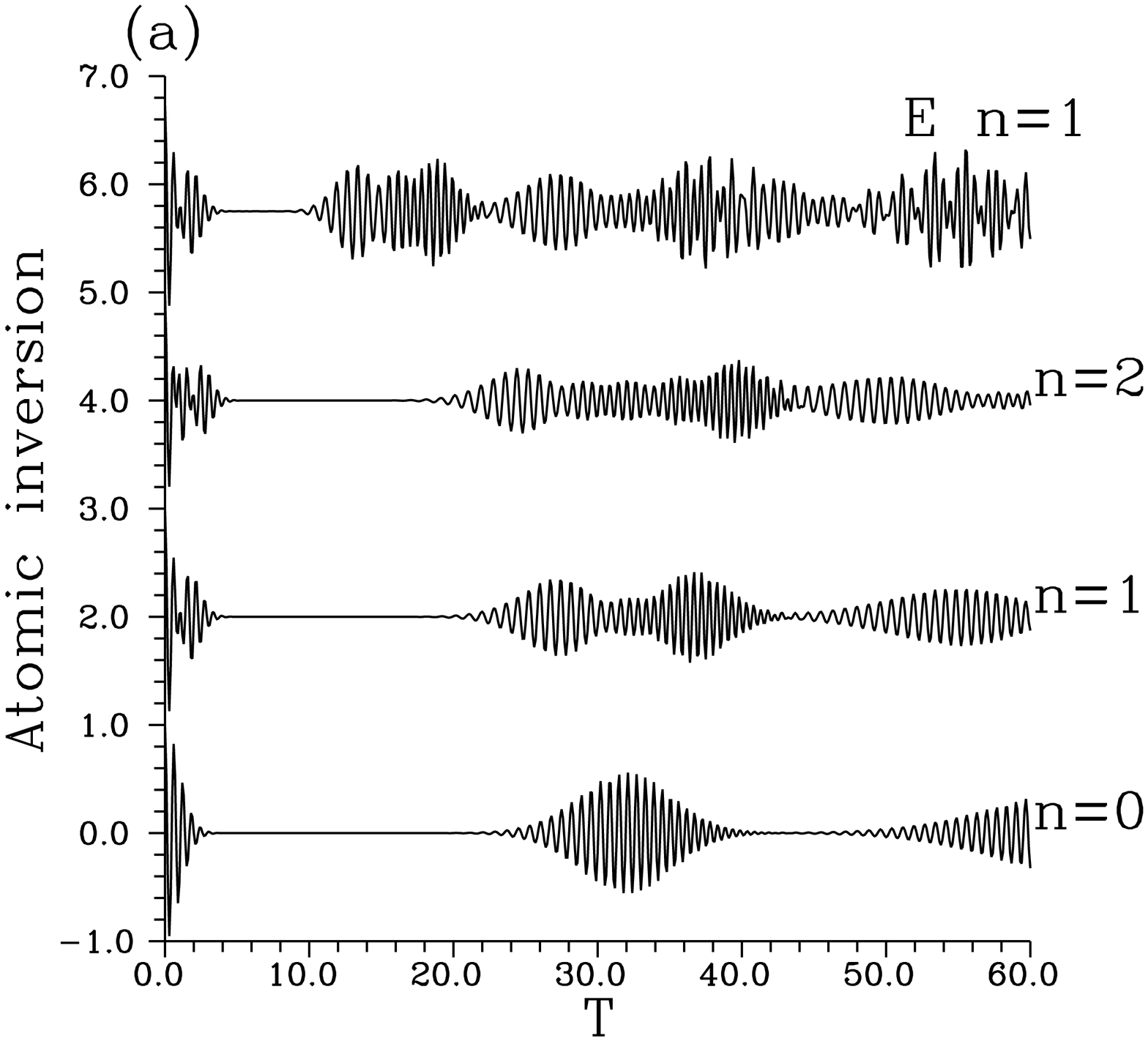}}
{\includegraphics[width=5cm]{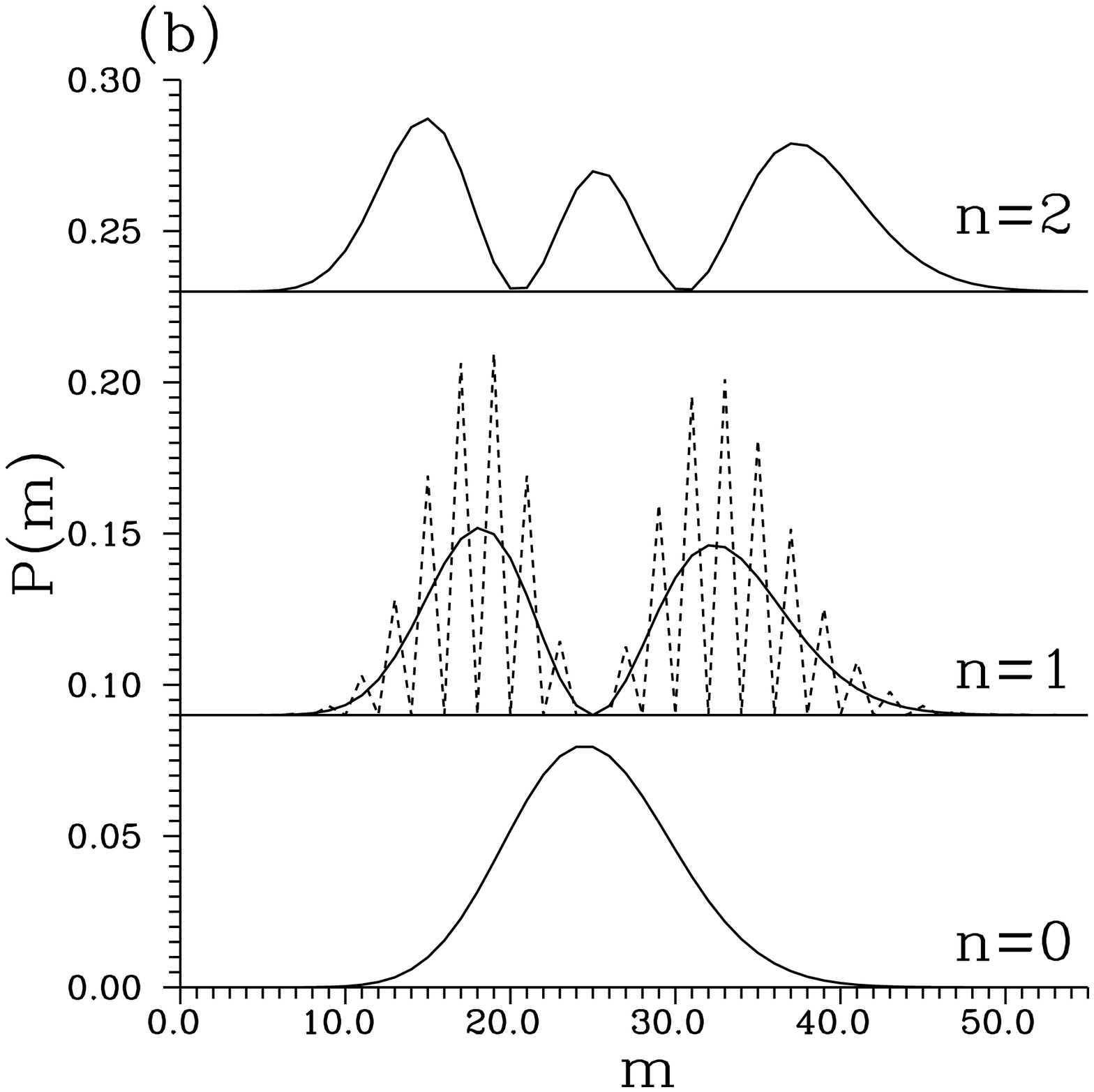}}
\caption{ The atomic
inversion (a) and the corresponding initial photon-number
distribution (b) for the Yurke-Stoler displaced number states and
for $\alpha=5 , k=1$; the different values of $n$ are specified in
the figures.  In Fig. 1a $E$ denotes that this curve is associated
with the even  displaced number states whose $P(m)$ is given in
Fig. 1b as dashed curve. Also  $\langle
\hat{\sigma}_{z}(T)\rangle$ and $P(m)$ are shifted from bottom to
top by $0,2,4,6$ and $0,0.09,0.23$, respectively.}
\end{figure}
Actually, for appropriate values of the parameters and when $\epsilon=i$,
 the connection between the peaks in $P(m)$  and the
 revivals in $\langle \hat{\sigma}_{z}(T)\rangle$ can be explained as follows \cite{rice}:
The different peaks in $P(m)$ have slightly different local mean photon
numbers $\bar{m}_{j}$ (see Fig. 1b) and thus each peak of $P(m)$ gives
its own revival-collapse pattern, which
interfere with that of the others to produce a very complicated
structure.
\begin{figure}
{\includegraphics[width=8cm]{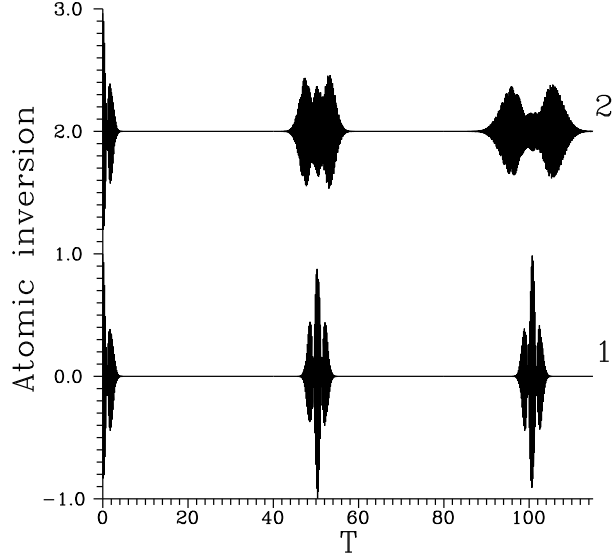}} \caption{ The  atomic
inversion $\langle \hat{\sigma}_{z}(T)\rangle$ for the asymptotic
(curve 1) and the exact  $2+ \langle \hat{\sigma}_{z}(T)\rangle$
(curve 2)  cases for even displaced number state with $\alpha=16,
k=1, r=0$ and $n=1$.}
\end{figure}
In this case  we have a sequence of the revivals with
slightly different periods $ T_{j}=\pi\sqrt{\bar{m}_{j}}, \qquad j=0,1,...$.
The echoes in the first revival will be separated by the intervals
$ (T_{j+1}-T_{j})$ whereas those in the second revival  will
be separated by $2 (T_{j+1}-T_{j})$ and so on.
Such a systematic behaviour depends upon the chosen values of the parameters,
in particular, $n$
(compare different curves in Fig. 1a). Furthermore, from this figure one
can observe that
the ringing revivals are remarkable  for the case $n= 2$.
For large values
of $n$ we noted that the chaotic behaviour is dominant and the collapse
time is shortened.
Now we turn our attention to the
superimposed case, i.e. the even or odd case (see curve E in Fig. 1a).
From this curve it is clear that  the collapse time
is smaller than that of the statistical mixture case owing to
 the interference in phase space.
 Precise information about this situation can  be obtained from
the asymptotic form for the atomic inversion, which can be evaluated via
the harmonic approximation technique  \cite{rice}.
For this purpose and for the sake of simplicity we confine ourselves to
 $|\alpha|>>1, k=1,n=1$ and $r=0$. In this case the argument
 of $\cos(.)$ in
(\ref{9})  can be expressed as \cite{rice}:

\begin{equation}
\sqrt{n+1}=\sqrt{\langle \hat{n}\rangle+n+1 - \langle \hat{n}\rangle} \simeq
\frac{1}{2}(\eta_{1}+\eta_{2}n), \label{16}
\end{equation}
where
\begin{equation}
\eta_{1}=
\sqrt{\langle \hat{n}\rangle} +\frac{1}{\sqrt{\langle \hat{n}\rangle}}
\qquad  \eta_{2}=\frac{1}{\sqrt{\langle \hat{n}\rangle}}, \label{13}
\end{equation}
and $\langle \hat{n}\rangle$ is the initial mean-photon number for the state
under consideration.
On substituting  (\ref{16}) into (\ref{9}) and after some minor algebra
we arrive at

\begin{eqnarray}
\begin{array}{lr} \langle
\hat{\sigma}_{z}(T)\rangle=\lambda_{\epsilon} \Bigl\{
(1+|\epsilon| ^{2}) f_{1}(T)\cos[T
(\eta_{1}+\eta_{2})+|\alpha|^{2}
\sin (\eta_{2}T)]\\
\\
+
2|\epsilon| f_{2}(T)\cos\phi \cos[T(\eta_{1}+\eta_{2})-|\alpha|^{2}
\sin (\eta_{2}T)]\Bigr\},
\label{14}
\end{array}
\end{eqnarray}
where
\begin{eqnarray}
\begin{array}{lr}
f_{1}(T)=\left[1-4
|\alpha|^{2}\sin ^{2}(\frac{\eta_{2}T}{2})\right]
\exp\left[-
2|\alpha|^{2}\sin ^{2}(\frac{\eta_{2}T}{2})\right],\\
\\
f_{2}(T)=\left[1-4
|\alpha|^{2}\cos ^{2}(\frac{\eta_{2}T}{2})\right]
\exp\left[-
2|\alpha|^{2}\cos ^{2}(\frac{\eta_{2}T}{2})\right].
\label{15}
\end{array}
\end{eqnarray}
\begin{figure}
  {\includegraphics[width=8cm]{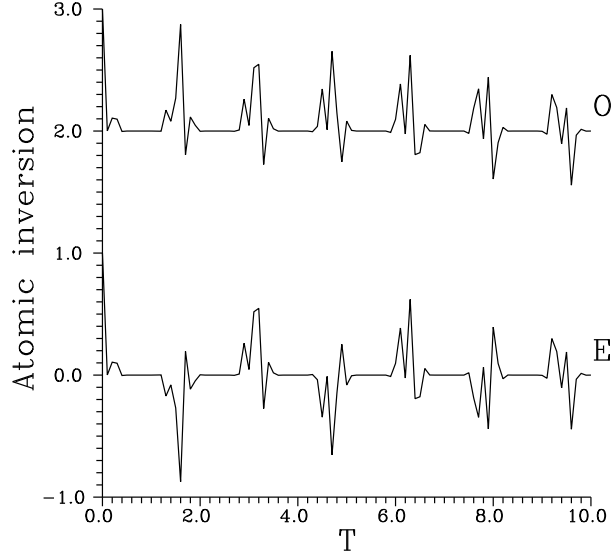}}
\caption{ The atomic inversion for even displaced Fock state
(E-curve-$\langle \hat{\sigma}_{z}(T)\rangle$) and odd displaced
Fock state (O-curve-$2+\langle \hat{\sigma}_{z}(T)\rangle$) for
$\alpha=5, \quad k=2,\quad n=1$  and $r=0$.}
\end{figure}
In (\ref{14}), $f_{1}(T)$ and $f_{2}(T)$ are the envelope functions
associated with
the statistical-mixture part (\ref{52})  and  the interference part
(\ref{53}) of the optical cavity
field density matrix, respectively.
These functions are periodic
with period $2\pi\sqrt{\langle \hat{n}\rangle}$ and their maxima occur
for  $T^{(S)}_{R}=2\pi\sqrt{\langle \hat{n}\rangle}$ and
$T^{(I)}_{R}=\pi\sqrt{\langle \hat{n}\rangle}$ where the superscripts $(S)$ and
$(I)$ mean that the times are associated with the statistical-mixture
 and interference parts, respectively, and the subscript $R$ denotes
the revival time.
It is obvious that $T^{(I)}_{R}=\frac{1}{2}T^{(S)}_{R}$, which
 is typical with that  of  the Schr\"{o}dinger-cat
states case \cite{vid}.
Further, the  term $1$ in the square bracket of (\ref{15}) corresponds to the
coherent component in the optical cavity, whereas
the second term is coming from the Fock state $|1\rangle$.
In other words,  the main revival associated with the
initial coherent light is
splitted into three parts showing  subsidiary revivals (echoes) in the atomic
inversion, e.g. for the statistical-mixture
part  splitting occurs at
$\sin (\frac{T}{2\sqrt{\langle \hat{n}\rangle}})=
\pm\frac{1}{2|\alpha|}$.
Furthermore, we have checked the case $n=2$ and
 found that, e.g., the statistical-mixture-part  envelope   splits into four parts controlled by the
equation
\begin{figure}

{\includegraphics[width=6cm]{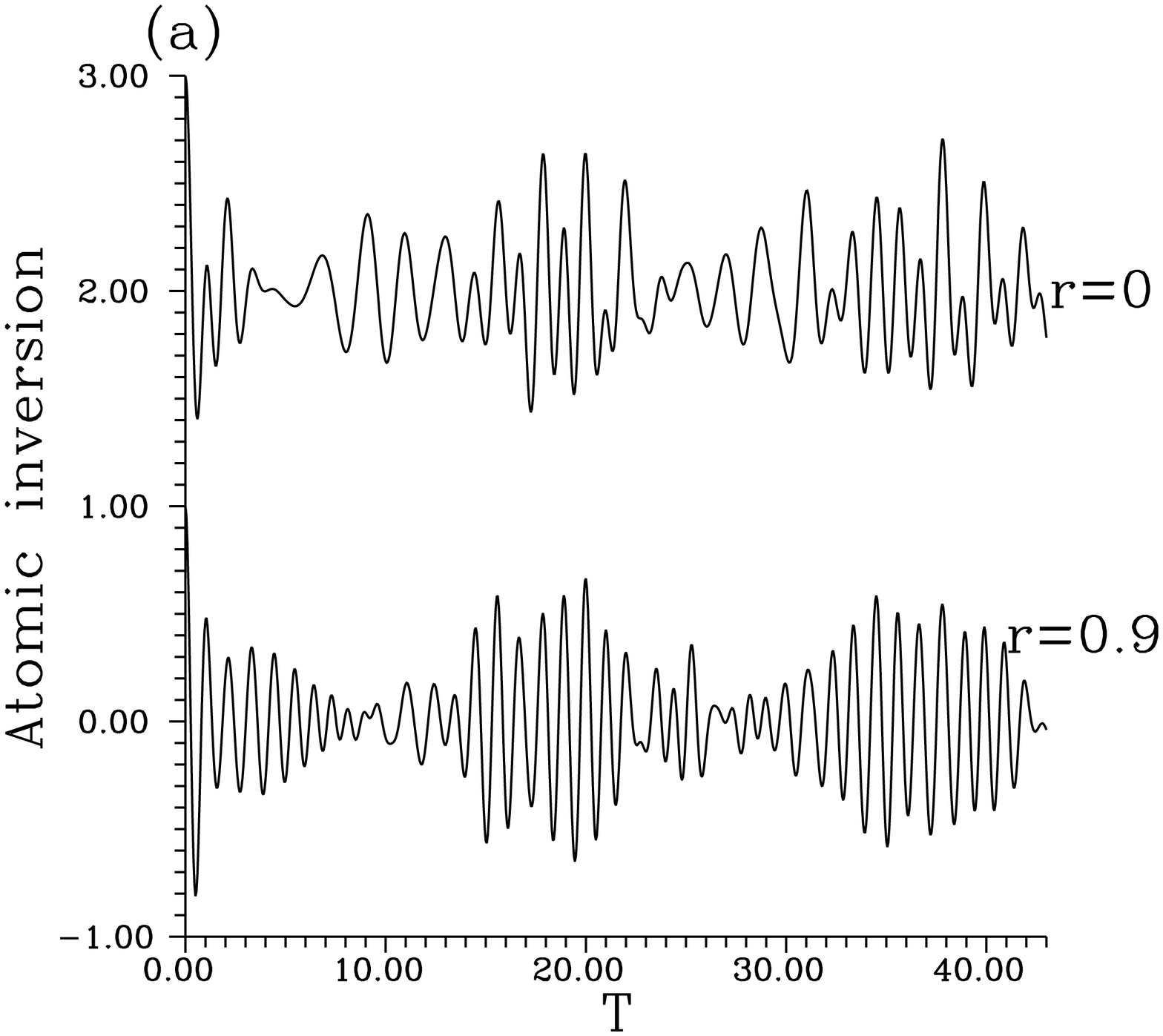}}
{\includegraphics[width=6cm]{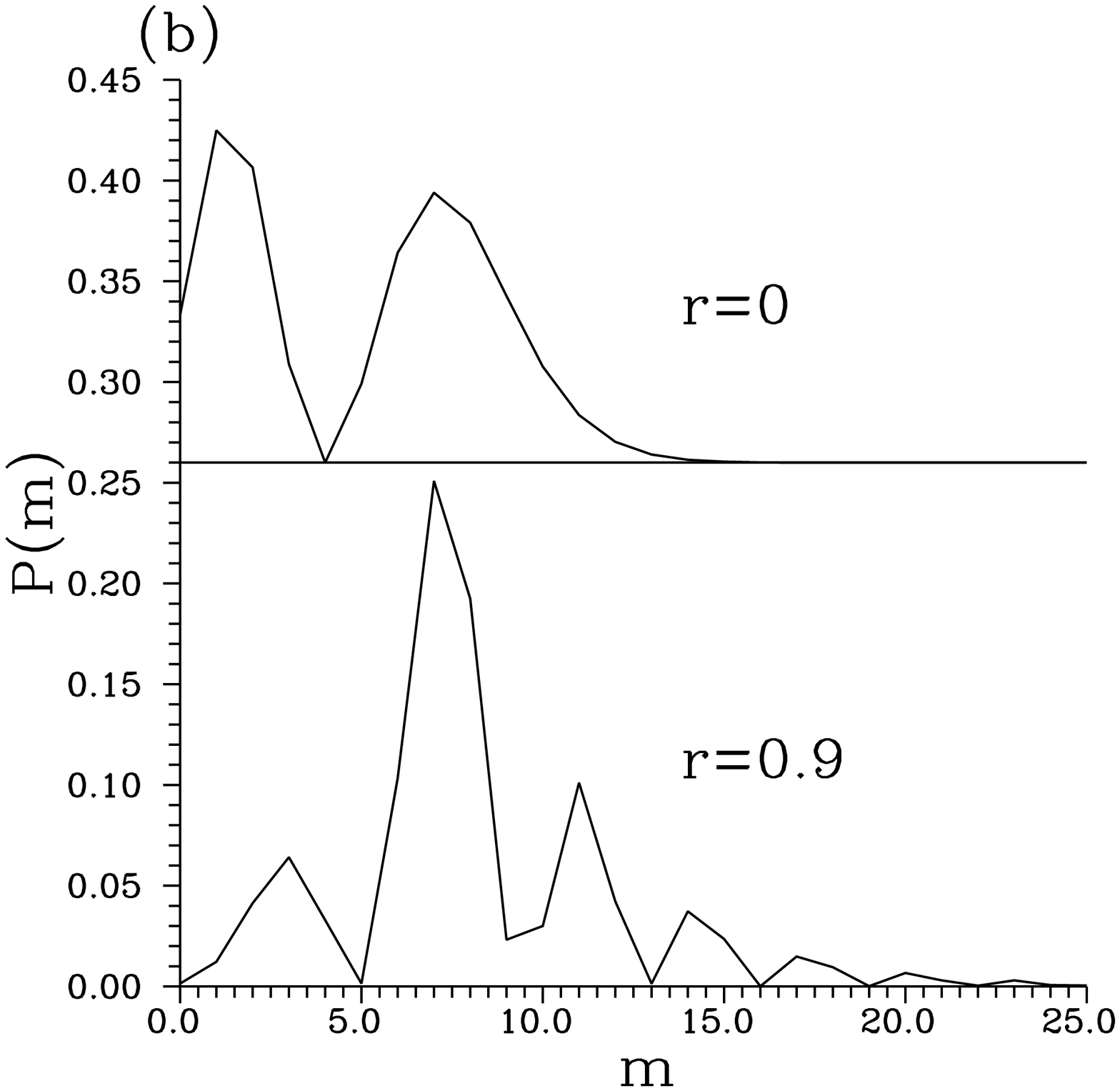}} \caption{ The atomic
inversion (a) and the corresponding initial photon-number
distribution (b) for $\alpha=3, \epsilon=i, \quad k=1,\quad n=1$
and different values of $r$ as indicated. $\langle
\hat{\sigma}_{z}(T)\rangle$ and $P(m)$ are shifted from bottom to
top by $0,2$ and $0, 0.26$, correspondingly.}
\end{figure}
\begin{equation}
8|\alpha|^{4}\sin^{4} (\frac{\eta_{2}T}{2})-8|\alpha|^{2}
\sin^{2} (\frac{\eta_{2}T}{2})+1=0.\label{15b}
\end{equation}
It is evident  that the Fock state $|2\rangle$ develops two additional terms in
 the asymptotic form (\ref{15b}), which provide more echoes in $\langle
 \hat{\sigma}_{z}(T)\rangle$ compared to that of the case $|1\rangle$.
In Fig. 2 we compare the behaviour of the asymptotic formula (\ref{14})
with that of the exact one  (\ref{9}) for the given
values of the parameters.  This figure reveals agreement and
disagreement between the exact and asymptotic behaviour.
Qualitatively, they are  approximately in agreement in  the
locations of the revival-collapse pattern as well as each revival
includes three portions, but
the corresponding revivals cannot provide typical  amplitude and
shape.  For the exact case
(curve $2$)  the revivals are broader  than those for the
asymptotic case.
This problem can be overcome by means of
the higher-order approximation in (\ref{16}) \cite{rice}, however, the
price that should be  paid, is that  we cannot obtain a simple closed form for
$\langle \hat{\sigma}_{z}(T)\rangle$.
On the other hand, excluding the initial
revival in Fig. 2, the origins of the first and second revivals,
respectively, are
statistical-mixture and interference parts of the initial density matrix
of the radiation field.
In conclusion the interference in phase space
decreases the collapse time and  the excitation number $n$  yields
echoes.

The issue we would like  to discuss here is that,
 qualitatively,
the nature of  quantum interference in phase space cannot manifest
itself in the behaviour of the atomic inversion
for the one-photon  JCM.
This is not the case for the multiphoton JCM where, e.g., the difference between
the even-state and odd-state cases becomes clear, in  particular, for $k=2$.
Actually, the two-photon JCM has been extensively studied and extended in many
direction (e.g. \cite{johia}).   Information about the case
$k=2$ is shown in Fig. 3 for  given values of the interaction parameters.
Clearly, the revivals in the two-photon case are much more compact than those
in the one-photon case.
Physically speaking  in the two-photon process the phase
correlation of the two superimposed optical cavities is transferred to the atomic
dynamics \cite{jos}. This is in contrast with the one-photon case where the atom absorbs
only one photon at a time. Furthermore, such behaviour can be
realized via the asymptotic form of this case, where we use the
asymptotic expansion  \cite{jos}
\begin{equation}
\sqrt{(n+1)(n+2)}\simeq \frac{3}{2}+n. \label{17a}
\end{equation}
Comparison between  (\ref{16}) and (\ref{17a}) leads to the fact
that the asymptotic results related to the two-photon case can be obtained from those
 of the one-photon case by simply setting $\eta_{1}=3$ and $\eta_{2}=2$.
  In this case the envelope functions revive periodically with period
$ T=\pi$ and then the revival time is $T=\pi/2$. So that the
collapse and revival times
 become approximately intensity independent and this
is responsible for such a systematic behaviour in the evolution of
the atomic inversion.
On the other hand, we have noted  that for $k\geq 3$ the revivals interfere to
produce chaotic population after the first collapse
\cite{maq}, i.e. the collapse-revival phenomenon of the atomic inversion
for multiphoton Jaynes-Cummings model is more dramatic.

We conclude this section by demonstrating that including squeezing in
the superimposed optical cavity can stimulate revival-collapse
pattern in the behaviour of $\langle \hat{\sigma}_{z}(T)\rangle$. In doing so,
 we plot Figs. 4 for the shown values of the parameters.
From Fig. 4b one can observe that for $r=0$,
$P(m)$ exhibits two peaks, which are smooth Gaussian peaks.
Such structure leads to  chaotic behaviour in $\langle \hat{\sigma}_{z}(T)\rangle$
(see Fig. 4a for the corresponding case) and can be explained as
follows: the second peak in $P(m)$ gives
 the familiar collapse-revival pattern in
 $\langle \hat{\sigma}_{z}(T)\rangle$, whereas the first peak
 (which is incomplete one) provides  oscillatory
 behaviour.
 Therefore, the competition between these two processes produces
such behaviour. On the other hand,
 when $r\neq 0$, i.e. squeezing in the optical cavity starts to play a role,
 one can observe that $P(m)$ has  oscillatory behaviour with
smooth Gaussian envelope, which causes the well-known revival-collapse
 phenomenon in $\langle \hat{\sigma}_{z}(T)\rangle$. This situation is similar to that of squeezed number states
 \cite{kim1}.

\section{Photon-number distribution and purity}
In this section we discuss the evolution of the
photon-number distribution $P(m,T)$ and  purity  $T_{f}$  for
the system under consideration.
As is well known photon-number distribution can be measured by photon detectors based on the
photoelectric effect.
Also, the characters of the field can be emphasized via
$P(m,T)$, which only at particular values of the interaction time resemble
 those of the initial pure states.
On the other hand, investigation of the purity for the present  quantum dynamical system
is important to know how far the atom and the field are  entangled.

We start the discussion by analyzing
$P(m,T)$, which  has the
form
\begin{equation}
P(m,T)=
P(m)\cos^{2} [T\sqrt{h(m,k)}]
+P(m-k)\sin ^{2} [T\sqrt{h(m-k,k)}], \label{10}
\end{equation}
where $P(m)$ are given by (\ref{5b}).
As the  optical cavity field starts to interact with the
atom the initial field  distribution   changes because of the
quantum dynamics described by the Hamiltonian (\ref{6}), however,
at specific times the field and the atom  subsystems return to their
initial states.    Basically such behaviour depends upon the values of
both the interaction time and the absorption parameter.
To be more specific, this  behaviour  is  established well for
$k=2$ and $4$, as we will show shortly. On the other hand,
for the standard JCM
at one-half of the revival time pure-superposition states can be generated
and the optical cavity field and the atom become completely disentangled
\cite{julio1,{buza},{hlad},{zahe}}.
These states are well established in
the strong-intensity regime, i.e. large values of $|\alpha|$.
Now  we discuss the possible  occurrence of such behaviour in  the present system.
To achieve our goal in a convenient way we restrict ourselves to the
case $\epsilon =0$ and $r=0$.
\begin{figure}
  {\includegraphics[width=6cm]{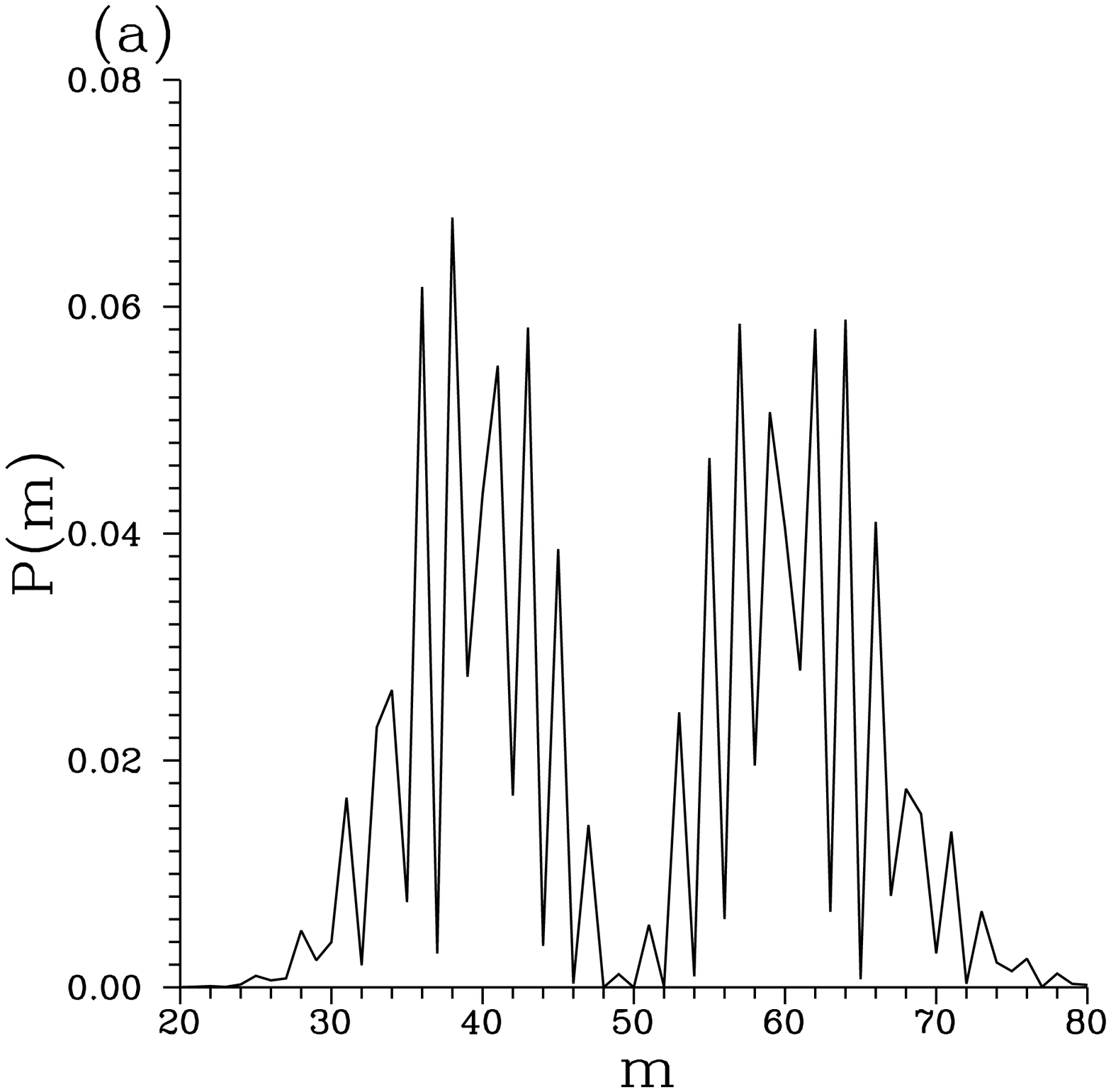}}
  {\includegraphics[width=6cm]{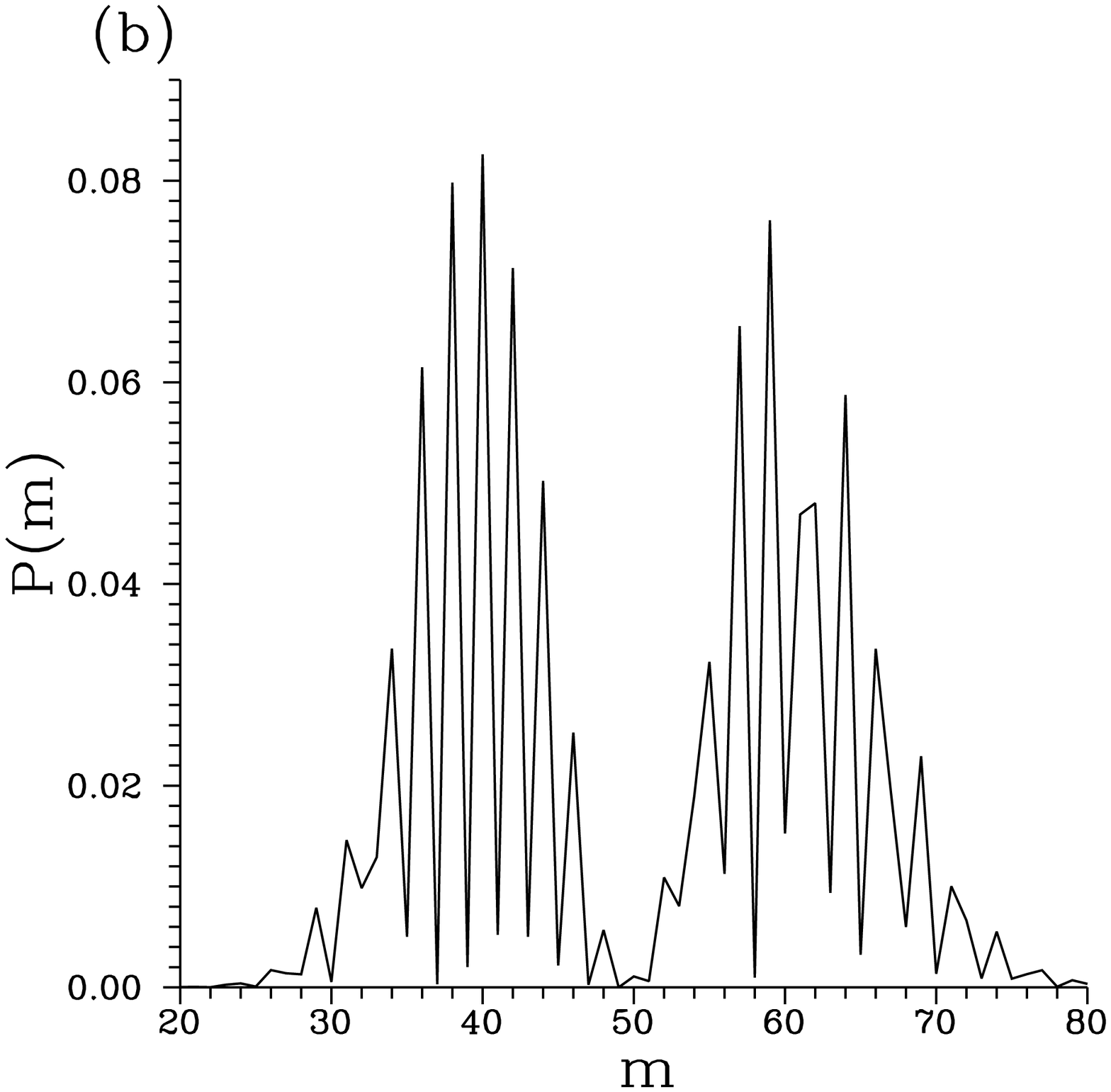}}
\caption{The photon-number distribution for $\epsilon=0,\alpha=7,
k=1$ and $n=1$ and for (a) $T=22.21$; (b) $T=20.64$. }
\end{figure}
In figures $5$ we plot $P(m,T)$
 for $T= \pi \sqrt{\langle\hat{n}(0)\rangle}=22.21$ (one-half of the revival time)
 (a) and $\pi \sqrt{\langle\hat{n}(0)\rangle}-0.5\pi=20.64$ (close to one-half of the
revival time) (b) and  for the given values of the parameters.
Actually, at these times the optical cavity field will be approximately
in  pure states this will be noticeable in the behaviour of the purity (see
Fig. 7 below). Furthermore,  Figs. 5 reveal that the optical cavity collapses
to superposition of displaced number states, particularly,
superposition of $ \hat{D}(\alpha) |1\rangle$.
This is obvious where $P(m,T)$ has oscillatory
behaviour  with two-Gaussian envelopes representative to such type of
states \cite{moya}.
Of course the oscillations in $P(m,T)$ are resulting from
the quantum interference in phase space.
The quantum properties for such type of states (,i.e., superposition of
displaced number states) have been
discussed in \cite{moya} and they have been
generated also in a state-selective measurement \cite{zeh1}.
We proceed, comparison of  Fig. 5a and Fig. 5b reveals that the
oscillations in $P(m,T)$ are extremely sensitive to the values of the interaction
time. Furthermore, comparison between the oscillatory behaviour in the
two peaks  in Fig. 5a (or in Fig. 5b) shows
that a clear asymmetry and this is connecting with the asymmetry in the
 initial photon-number
distribution of the states $\hat{D}(\alpha) |1\rangle$ (see Fig. 1b).
Actually, by changing the types of the initial  state we can obtain
different forms of cat states.

Now we discuss the case $k=2$ (two-photon absorption) whose scaled revival time
is $T_{R}=\pi$, where we still consider  $\epsilon=0$. We found that the optical cavity field at $T=sT_{R}$
where $s$ is positive integer, collapses to pure state typical
(or similar to) the initial states, while the atom becomes either in
the ground state or in the excited state.
In other words, the atom and the field subsystems become completely
disentangled at these specified times.
Such behaviour  can be
analytically determined provided that $\alpha$ is large and $n$ is
finite through the substitution of both $T=T_{R}=\pi$ and
 (\ref{17a}) into the state vector obtained from (\ref{8}),
 we arrive at:

\begin{eqnarray}
\begin{array}{lr} |\psi(T=T_{R})\rangle
=\sum\limits_{m=0}^{\infty}C_{m}(\alpha,r,n) \Bigl\{\cos
[T_{R}(m+\frac{3}{2})]|m,+\rangle- i
\sin [T_{R}(m+\frac{3}{2})]|m+2,-\rangle\Bigr\} \\
 \\
=|\psi^{(2)}_{f}(0)\rangle\bigotimes \exp(\frac{i}{2}\pi)|-\rangle, \label{18a}
\end{array}
 \end{eqnarray}
where $|\psi^{(2)}_{f}(0)\rangle$ denotes the field state
having the form
\begin{equation}
|\psi^{(2)}_{f}(0)\rangle=\sum\limits_{m=0}^{\infty}C_{m}(\alpha,r,n)\exp (im\pi) |m+2\rangle.
 \label{18b}
 \end{equation}
Expression (\ref{18b}) shows that
 the cavity field possesses photon distribution  as that
of the initial one, but with two-photon shift.   On the other hand,
by applying the above   procedures   for the case
$T=2T_{R}=2\pi$, one can easily show that the system returns back
to its initial form, i.e. the
atom is in its excited state and the field  is in
$|\psi^{(1)}_{f}(0)\rangle$ given by (\ref{1}) with $\epsilon=0$.
Generally, we conclude that the system evolves periodically according to
the following relation:

\begin{equation}
|\psi^{(1)}_{f}(T)\rangle=\left\{
\begin{array}{rl}
|\psi^{(2)}_{f}(0)\rangle\;\;&{\rm for}\;T=sT_{R}\;\; {\rm and}\;\; s=1,3,5,..,\\
|\psi^{(1)}_{f}(0)\rangle\;\;&{\rm for}\;T=sT_{R}\;\;
{\rm and}\;\; s=2,4,6,..,.
 \end{array}
\right.
\label{18c}
\end{equation}
Expression (\ref{18c}) can be understood as follows: when the interaction is going on the atom decays (spontaneously and
stimulately) from the excited state to the ground  state contributing  its photon
to the cavity field and the system becomes disentangled
during interaction time $T=T_{R}$. This means that  the atomic excitation energy is transferred to the
field \cite{hlad}, i.e. an information
about the initial state of the atom is stored in the field.
 Furthermore, when the interaction continues the feed back between
field and atom occurs and after $T=2T_{R}$ the system returns back to its
initial form.
This behaviour is also  visible in the evolution of the atomic inversion
and reflects the periodic nature of the system.
We should stress that the origin of such behaviour is
two-fold: (i) We have considered the system is
completely isolated (the interaction with the environment is neglected).
(ii) We treat the system  in the RWA, which ensures that whenever $k$
photons are lost from the field the atomic state must change from ground
to excited states, or vice versa. Furthermore, there is a similarity
between  the present behaviour and  that of the frequency
conversion device in which the energy interchanges  periodically  between
the signal and idler modes.
On the other hand, it has been shown  that for the case
$k=2$ and at the quarter of the revival time  superposition states
can be generated  \cite{hlad}.
\begin{figure}
  {\includegraphics[width=8cm]{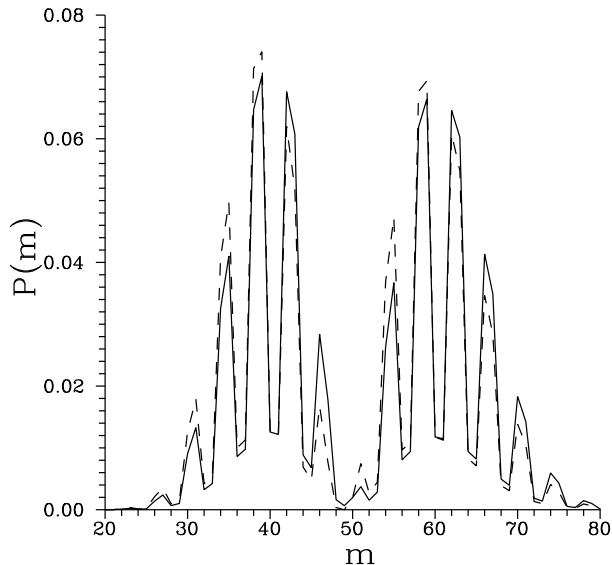}}
\caption{The photon-number distribution for $k=2$,
$T=\frac{1}{4}\pi$ and the other parameters as those in Fig. 5.}
\end{figure}
Presently we show analytically using a similar  technique as that in
\cite{julio1} that  similar behaviour can occur.
We restrict ourselves to the case $\epsilon=0, n=1$ and $\alpha$ is very
large. On
substituting  $T=\frac{1}{4}T_{R}=\frac{1}{4}\pi$ and (\ref{17a}) into the
state vector  (\ref{8}), we obtain

\begin{eqnarray}
\begin{array}{lr} |\psi(T=\frac{1}{4}T_{R})\rangle=
\sum\limits_{m=0}^{\infty} C_{m}(\alpha,r,n)\Bigl\{\cos
[\frac{1}{4}T_{R}(m+\frac{3}{2})]|m,+\rangle
- i\sin [\frac{1}{4}T_{R}(m+\frac{3}{2})]|m+2,-\rangle \Bigr\}\\
 \\
=
\sum\limits_{m=0}^{\infty}
\cos [\frac{1}{4}T_{R}(m+\frac{3}{2})]
\Bigl\{C_{m}(\alpha,r,n) |+\rangle +iC_{m-2}(\alpha,r,n) |-\rangle\Bigr\}
|m\rangle.
\label{19a}
\end{array}
 \end{eqnarray}
For the case $n=1$ the relation between $C_{m}(\alpha,r,n)$
and $C_{m-2}(\alpha,r,n)$ reads
\begin{equation}
C_{m-2}(\alpha,r,n)=\sqrt{\frac{m^{2}}{\alpha^{4}}-\frac{m}{\alpha^{4}}
}\left[\frac{m-\alpha^{2}-2}{m-\alpha^{2}}\right]
C_{m}(\alpha,r,n).  \label{19b}
 \end{equation}
It is worth remembering that the initial photon distribution has two-peak
structure (see Fig. 1b) each of which has a Poissonian envelope.
Further, for very large values of $\alpha$ one can find out that  the terms contribute effectively to
the summation in (\ref{19a}) are those for which the values of $\alpha^{2}$ and
$m$ are comparable.  Therefore, the first term in the square root of
(\ref{19b}) tends to unity while the second one tends to zero.
Also the terms in the square bracket reduces to unity where $\alpha^{2}>>2$.
Considering all these suggestions, (\ref{19a}) can be rewritten as
\begin{figure}
  {\includegraphics[width=6cm]{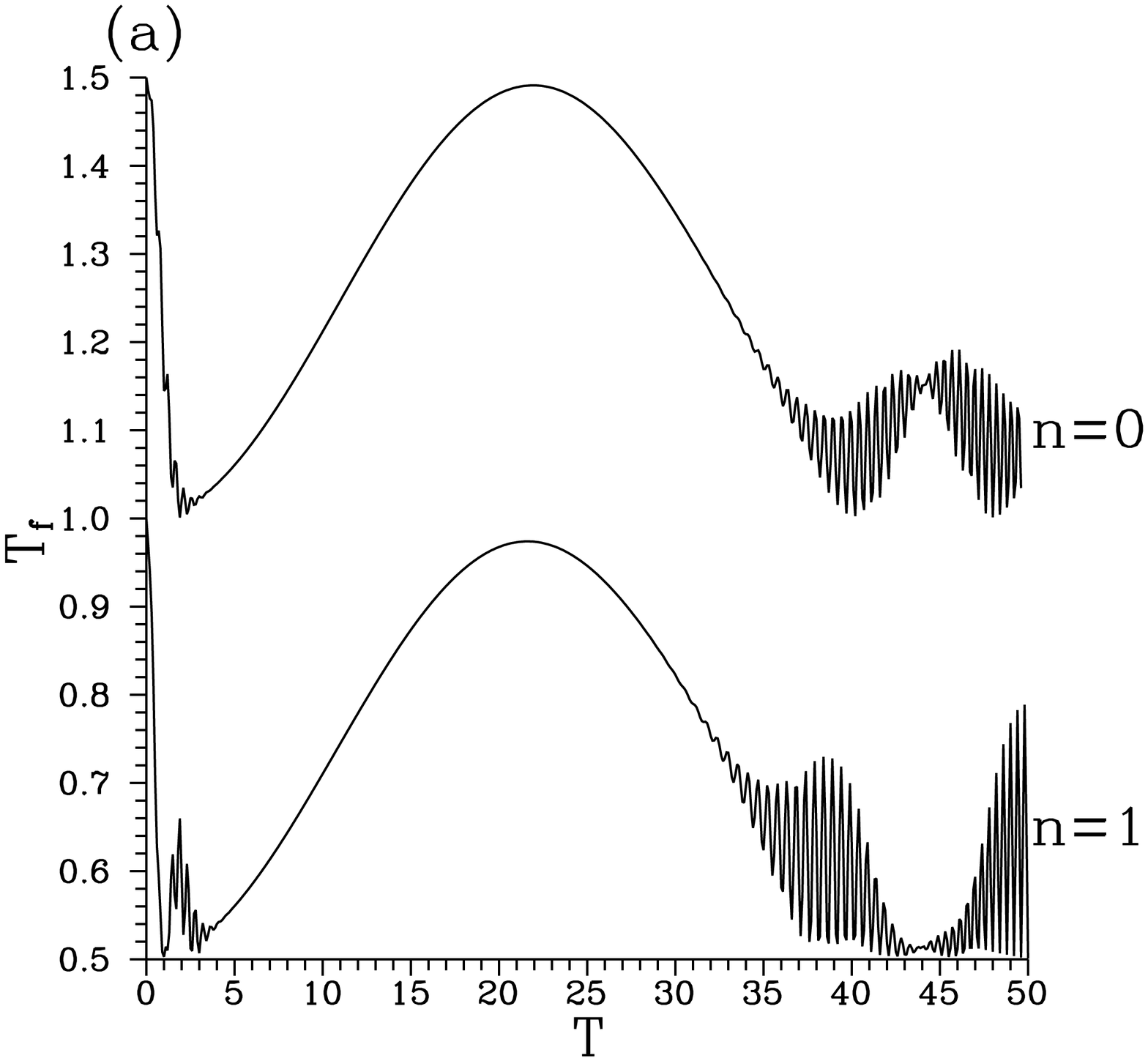}}
  {\includegraphics[width=6cm]{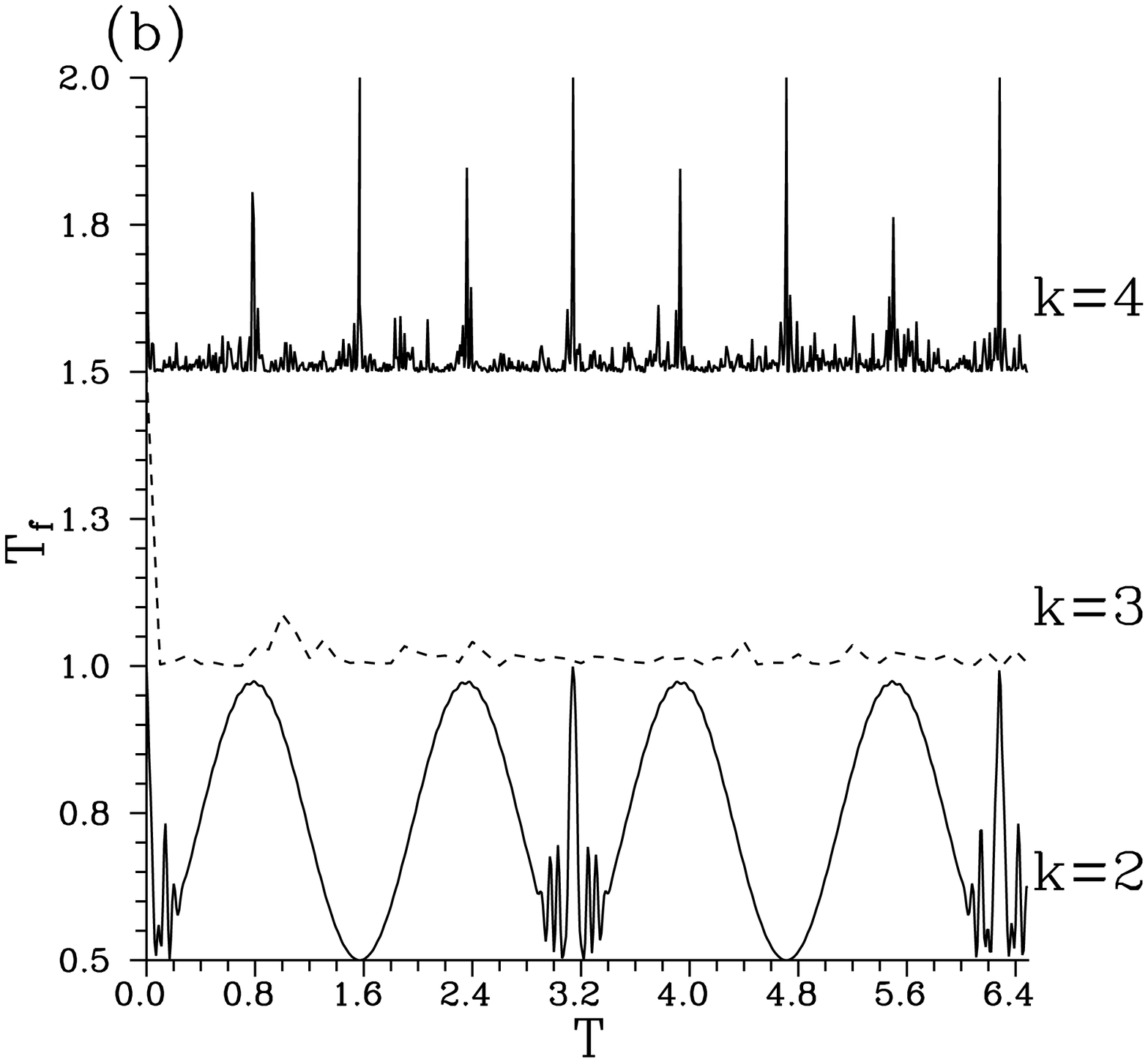}}
  \caption{The purity against the scaled time $T$
for: a) $(\alpha,k,\epsilon)=(7,1,0)$ ; and b)
$(\alpha,n,\epsilon)=(7,1,0)$. In (a) for $n=0$ we plot $T_{f}+0.5$.
In (b) for $k=3$ and $4$ we plot $T_{f}+0.5$ and $T_{f}+1$, respectively.}
\end{figure}

\begin{equation}
|\psi(T=\frac{1}{4}T_{R})\rangle= \sum\limits_{m}^{\infty}
C_{m}(\alpha,r,n)\cos [\frac{1}{4}T_{R}(m+\frac{3}{2})]|m\rangle
\bigotimes\Bigl\{ |+\rangle+i |-\rangle\Bigr\}.
 \label{19c}
 \end{equation}
From this  expression  it is clear  that at the quarter of the revival time
the atom and the field become asymptotically disentangled
and  the atom evolves in an atomic superposition  state, whereas the cavity field
collapses  to  superposition of  states whose forms
 are somewhat similar to those given in
 \cite{yur1}.  From (\ref{19c}) it is clear  that
$\cos [\frac{1}{4}T_{R}(m+\frac{3}{2})]\neq 0$ for $m$ integer
 meaning that the $P(m,T)$  has oscillatory behaviour but
cannot have perfect oscillations (,i.e., $P(m,T=T_{R}/4)=0$) more than
those included in the initial distribution of the  optical cavity.
This behaviour is clear in Fig. 6 where we plot the $P(m,T)$ at
$T=T_{R}/4$ for the same situation as in Fig. 5 but $k=2$. Furthermore,
in Fig. 6 the solid and dashed curves are associated with the exact
(\ref{10}) and asymptotic (\ref{19c}) forms of $P(m,T)$.
From Fig. 6 one can observe that
the solid and dashed curves include significantly similar
behaviour. In other words, under these conditions
the asymptotic  expression (\ref{19c}) can be used to  describe
 the optical cavity field well.
It is worth mentioning that  approximate perfect oscillations
can be achieved in the behaviour of this case if one uses the replacement
 $T_{R}\rightarrow T_{R}\pm 0.08$, however, the behaviour of
$P(m,T)$ will be very close to that in Fig. 5b. Furthermore, in (\ref{6}) if
$\hat{a}^{k}$ replaced by $\hat{a}\sqrt{\hat{a}^{\dagger}\hat{a}}$
(i.e., the system becomes intensity-dependent JCM) the system can generate
perfectly even and odd superposition of displaced number states
(considering the initial state is the displaced Fock state) according
to  the atom is prepared initially in the ground and in the excited states,
 respectively.

We close this section by analyzing the purity for the optical cavity field via
the relation $T_{f}={\rm Tr}[\hat{\rho}_{f}^{2}(T)]$ where
$\hat{\rho}_{f}(T)
={\rm Tr}_{a}[\hat{\rho}(T)]$ is the field reduced density matrix,
 whereas $\hat{\rho}(T)$ is the total density matrix of the system (\ref{8}).
When $T_{f}=1$ the reduced density matrix describes pure state, however,
for $T_{f}<1$  the field will be in a statistical
mixture state. The smaller the purity is, the less pure
the state will be.
Further, when $T_{f}=0.5$ the field and the atom will be in a maximally entangled state.
As we mentioned above $T_{f}$ gives information about  entanglement
of the components of the system, which forms the basis of
experiments in the realm quantum information. Further,  the
disentanglement of the two quantum systems suggests interesting
applications, e.g., in atomic states preparation through interacting quantum
systems or in the choice of optimum time of flight in a micromaser to
determine the cavity field \cite{phon}.  Moreover, from the point of view of the quantum theory of
measurement, the decoupled field state can be regarded as a pointeer
basis since all the information about the atomic state has been
transferred to the field \cite{julio1}.
We proceed, in Figs. 7 we plot the evolution of the $T_{f}$
for the given values of the interaction parameters. Comparison between
the curves associated with $n=1$ and  $n=0$ in Fig. 7a (for fixed value of $k$)
 shows that including Fock state $|1\rangle$ in the coherent optical cavity
give rises to the oscillatory behaviour in $T_{f}$, changes the
parity and achieve maximal entanglement at
$T=2\pi\sqrt{\langle\hat{n}(0)\rangle}$ (revival time).
Actually, the oscillatory behaviour in $T_{f}$ occurs in a good
correspondence  with the revival in $\langle \hat{\sigma}_{z}(T)\rangle $.
Further, for both cases ($n=0,1$) the field collapses to a pure state at the middle of the
collapse time ($\pi\sqrt{\langle\hat{n}(0)\rangle}$).
Also the figure does not show a well-defined ringing revival for the
case $n=1$, which is in contrast with the behaviour of the
 $\langle \hat{\sigma}_{z}(T)\rangle $.
Furthermore, we have noted that increasing the value of $n$ will not give significant
change in the overall behaviour of $T_{f}$ except increasing the oscillations.
However, for the superimposed optical cavity, we have noted also that
the state of the field never gets as pure as in the Fig. 7a and the system
remains in a maximally (or partially) entangled state most of the time.
On the other hand, Fig. 7b is given for  fixed value of $n$ and different values of
$k$ as indicated. For the case $k=2$,
$T_{f}$ evolves smoothly and the field collapses to an approximate pure state
at $T=\pi/4$ and $3\pi/4$,  which is long lived as compared to those
occurring at
times  multiples of $\pi$. As we mentioned above for $T=s\pi,
s=0,1,2..,$  the optical cavity field returns
 to its initial form.
Furthermore, the field and the atom are long lived  maximally entangled  at
$T=\pi/2$ and $3\pi/2$.
It is evident that $T_{f}$ has a systematic behaviour as
 $\langle \hat{\sigma}_{z}(T)\rangle $.
Now we turn our attention to the case $k=4$. One can observe that
$T_{f}$ displays
pure states instantaneously with period $\pi/2$, however, at
$T=\pi/4$ and $3\pi/4$, the field will be in a partially mixed state. Apart
from these values of interaction time
the field and the atom are approximately maximally entangled.
The similarity and dissimilarity between the $T_{f}$
for the cases $k=2$ and $k=4$ are remarkable.
On the other hand, for the case $k=3$, the field goes to  approximate
mixture states just after the interaction is turned on
meaning  that the field and the atom almost retain a strong
entanglement  all over interaction times.

In conclusion,  the behaviour of $P(m,T)$ and $T_{f}$ are in a good
agreement with each others.
Also, the analysis of the purity gives information that could not be predicted
by looking at the inversion or the photon-number distribution solely.
We have not considered the influence of squeezing parameter in the
present section since it generally increases  the oscillatory behaviour
in the $P(m,T)$.
\section{Sub-Poissonian statistics and quadrature squeezing}
In this section we discuss the sub-Poissonian statistics  and quadrature squeezing for the
system under consideration.  Both of these quantities can be used as a
measure for  the nonclassical effects.
Firstly, the sub-Poissonian statistics can be analyzed using
 Mandel  $Q$ parameter \cite{mol1} having the form
\begin{equation}
Q(T) =\frac{\langle \left(\triangle\hat{n}(T)\right)^{2}\rangle -
\langle \hat{n}(T)\rangle }
{\langle \hat{n}(T)\rangle}.  \label{20a}
\end{equation}
Formula (\ref{20a}) characterizes the deviation from Poissonian
statistics.
 It holds that $-1\leq Q(T)<0$ for sub-Poissonian statistics
(nonclassical effect), $Q(T)>0$ for super-Poissonian statistics
(classical effect) and  $Q(T)
=0$  for Poissonian statistics  (the standard one).
Sub-Poissonian light can be measured by means of a set of photodetectors.
Further, this  light has several applications, e.g. in the gravitational
 wave detector and quantum nondemolition measurement, and can be
 generated  in semiconductor
lasers \cite{yama} and in the microwave region using masers operating in the
microscopic regime \cite{rem}.

Secondly, in order to investigate squeezing we define two quadrature operators as
$\hat{X}=\frac{1}{2}[\hat{a}+\hat{a}^{\dagger}], \quad
\hat{Y}=\frac{1}{2i}[\hat{a}-\hat{a}^{\dagger}]$.  These quadratures
 relate to the conjugate electric and magnetic field operators
 of the electromagnetic wave and satisfy the commutation rule
$[ \hat{X},\hat{Y}] =\frac{i}{2}$.
Therefore,  the uncertainty relation reads
$\langle (\triangle\hat{X}(T))^{2}\rangle
  \langle (\triangle\hat{Y}(T))^{2}\rangle \geq \frac{1}{16}$.
So  we can say that the system is able to produce squeezing
if the squeezing factor $F(T)=4\langle (\triangle \hat{X}(T))^{2}\rangle-1<0$  or
$S(T)=4\langle (\triangle \hat{Y}(T))^{2}\rangle-1<0$.
As is well known squeezed light can be measured by a homodyne detection in
which the signal is superimposed on a strong coherent beam of the local
oscillator.
Furthermore, quite recently it has been shown experimentally that there
is an evidence of  squeezed light in the biological systems \cite{popp}.

The required  expectation values to treat these quantities can be calculated
using  the following
relation
\begin{equation}
\langle
\hat{a}^{\dagger s}(T)
\hat{a}^{s'}(T)\rangle={\rm Tr}\left[\hat{\rho}(T)
\hat{a}^{\dagger s}(0)\hat{a}^{s'}(0)\right],  \label{20b}
\end{equation}
where $\hat{\rho}(T)$ is given by (\ref{8}) and $s,s'$ are positive
integers.
\begin{figure}
  {\includegraphics[width=6cm]{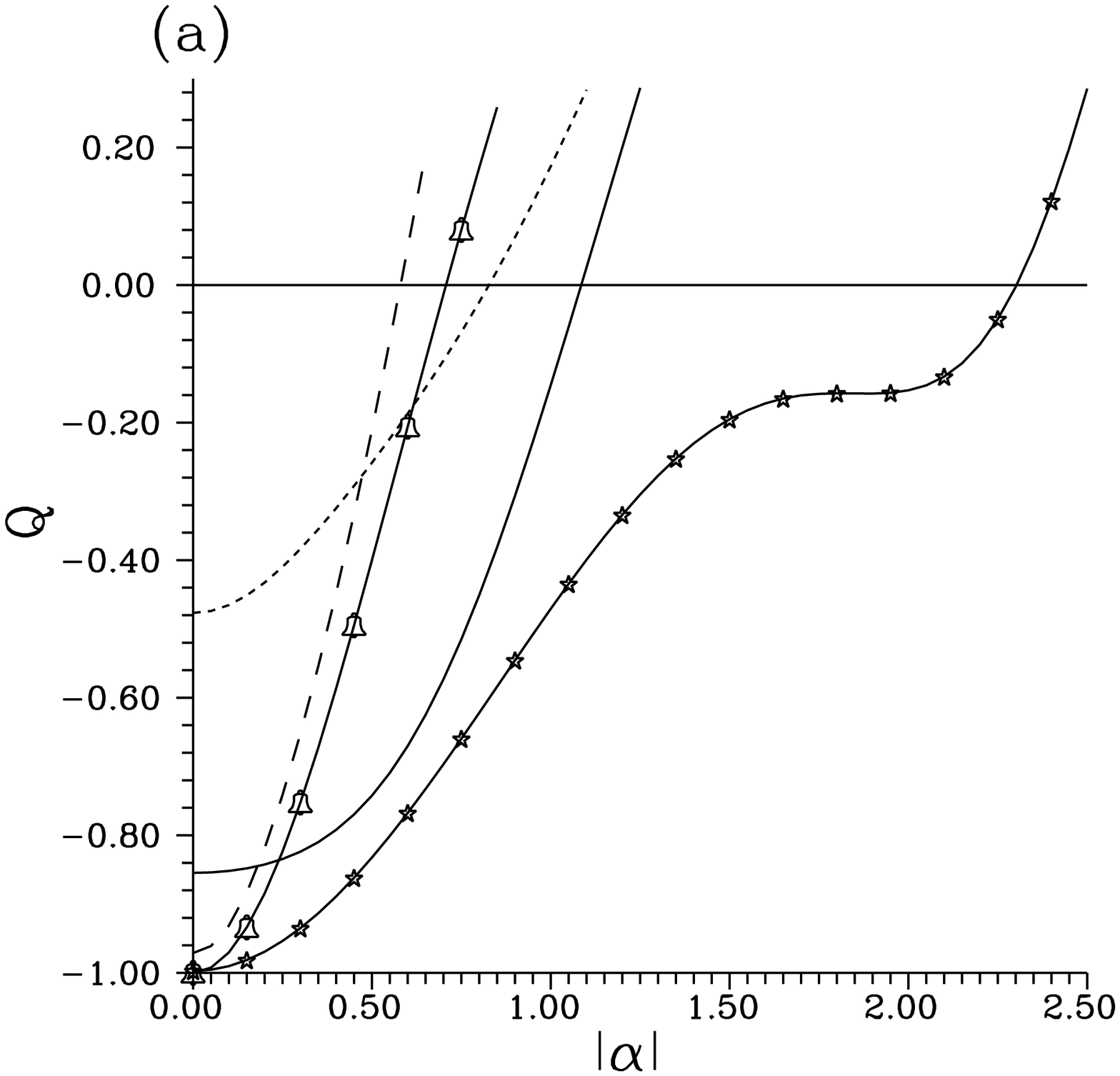}}
  {\includegraphics[width=6cm]{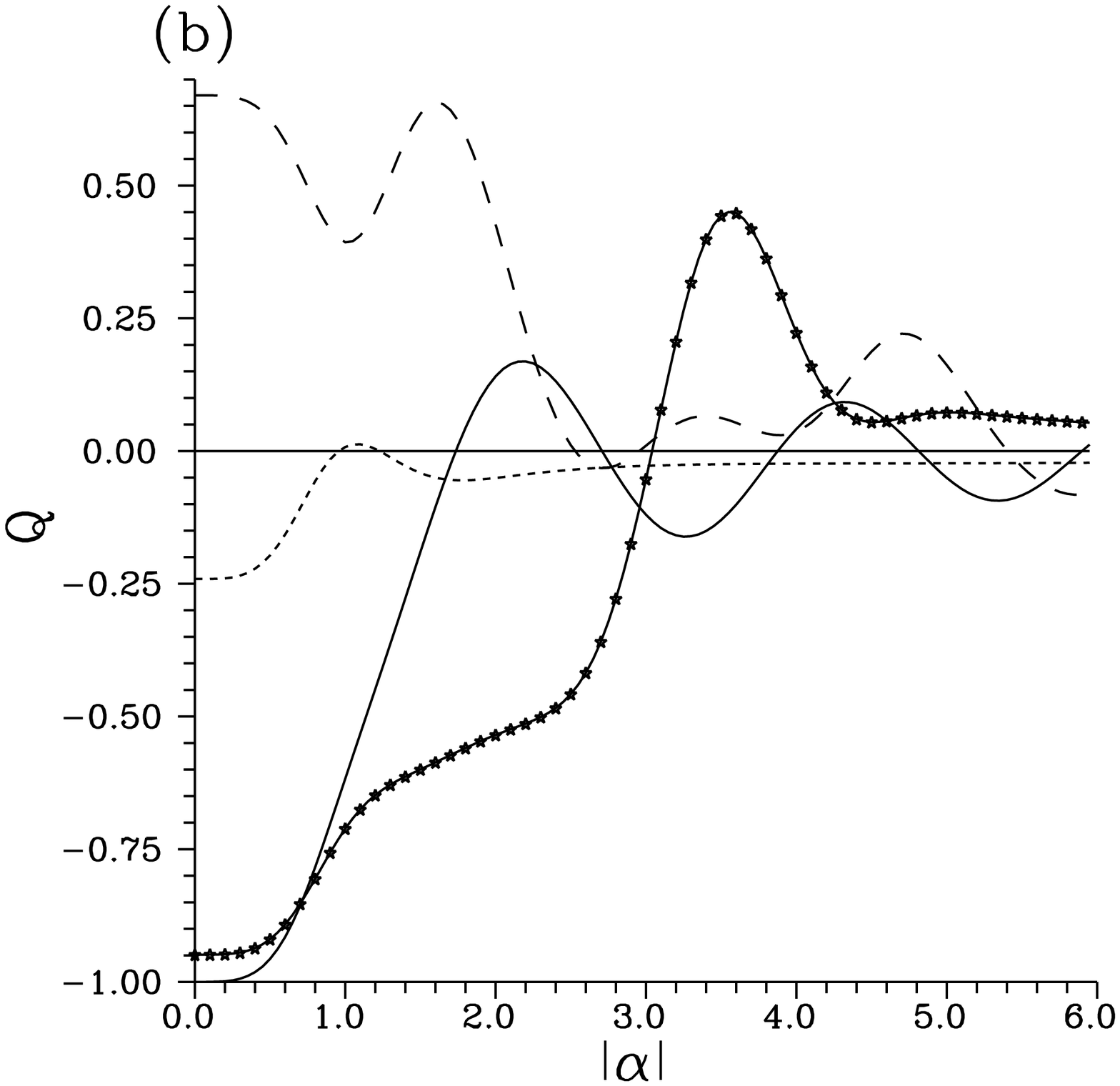}}
\caption{Mandel $Q$ parameter against $|\alpha|$ when the optical
cavity field is initially prepared in $\hat{D}(\alpha)|1\rangle$
 (a) and
even coherent state (b)
for $T=1.578$ and for
 $k=1,2,3,4$ corresponding
to solid, short-dashed, long-dashed and star-centered curves.
In (a)  the bell-centered curve  is for zero-interaction time.
Straight line is the sub-Poissonian  bound.}
\end{figure}
Before discussing the sub-Poissonian statistics for the present system
it is worth reminding that  for the standard JCM the sub-Poissonian
statistics have been realized
 \cite{kim2}. Furthermore, the establishment  of such statistics
 for input squeezed coherent  states relies on the
competition between the squeezing parameter $r$ and coherent amplitudes
$|\alpha|$ \cite{maq}. In other words, as the values of $r$ increase
the amount of
sub-Poissonian statistics (resulting from the coherent component)  is decreased
and eventually disappeared.
For the system under consideration the squeezing parameter
plays the same role, i.e. it destroys the sub-Poissonian statistics.
In general, we found that $Q(T)$ exhibits
collapse-revival pattern as well as ringing revivals, which reflects the time evolution of the atomic
inversion. In this case the photon statistics already oscillates between
sub-Poissonian and super-Poissonian  statistics based on the values of the
interaction parameters.
Here  we would like to address two facts:
 (i) The system can enhance the sub-Poissonian interval for the initial
 sub-Poissonian states.
(ii) The system can generate  sub-Poissonian light for  an initial
non-sub-Poissonian superimposed state (e.g. even coherent states).
Information about these facts is included in Figs. 8a and b for given values
of the parameters.
In these figures we have plotted Mandel $Q$ parameter
in the stationary regime at specific value of interaction time,
since the time-dependent coefficients in the present system
are periodic in time. The  selected value of the interaction time
gives maximum sub-Poissonian statistics for the case $k=1$ when the
optical cavity field is initially prepared in the even coherent states.
From Fig. 8a, one can observe
that the curves are smooth and the most sub-Poissonian statistics
 occur for $k=3$ and $4$. Also this figure gives the coherent
amplitude  interval
over which the sub-Poissonian  light  can be obtained for different values
of the absorption parameter. It is evident that the intervals for the cases
$k=1,2,4$ are greater than those for the initial one
(compare different curves with the bell-centered one).
This shows that how one can enhance the sub-Poissonian interval based on
the value of the absorption parameter.
Furthermore,  the amount of
sub-Poissonian statistics decreases almost exponentially as the magnitude
$|\alpha|$ increases similar to  that of the standard JCM \cite{hill}.
\begin{figure}
{\includegraphics[width=5cm]{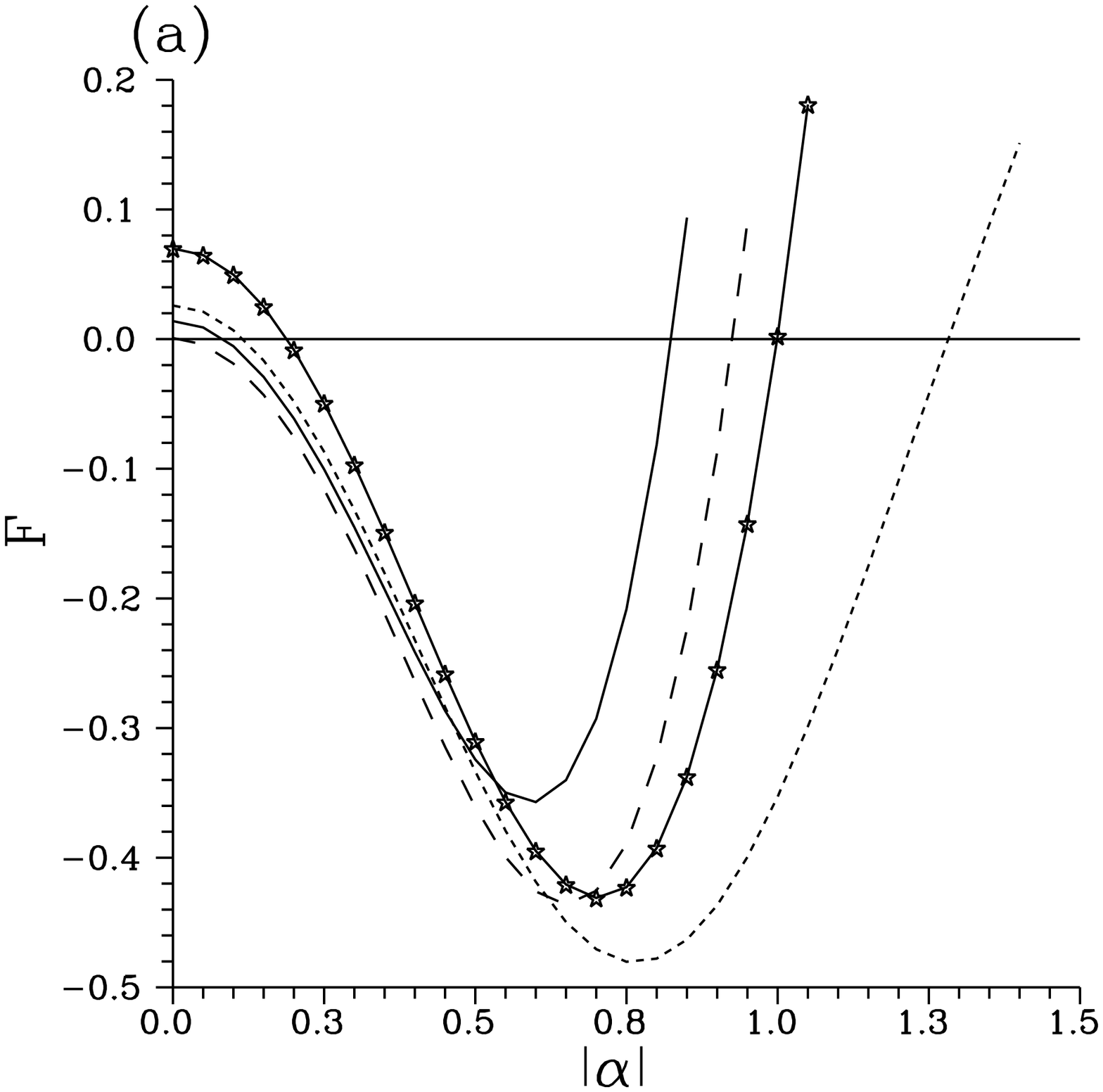}}
{\includegraphics[width=5cm]{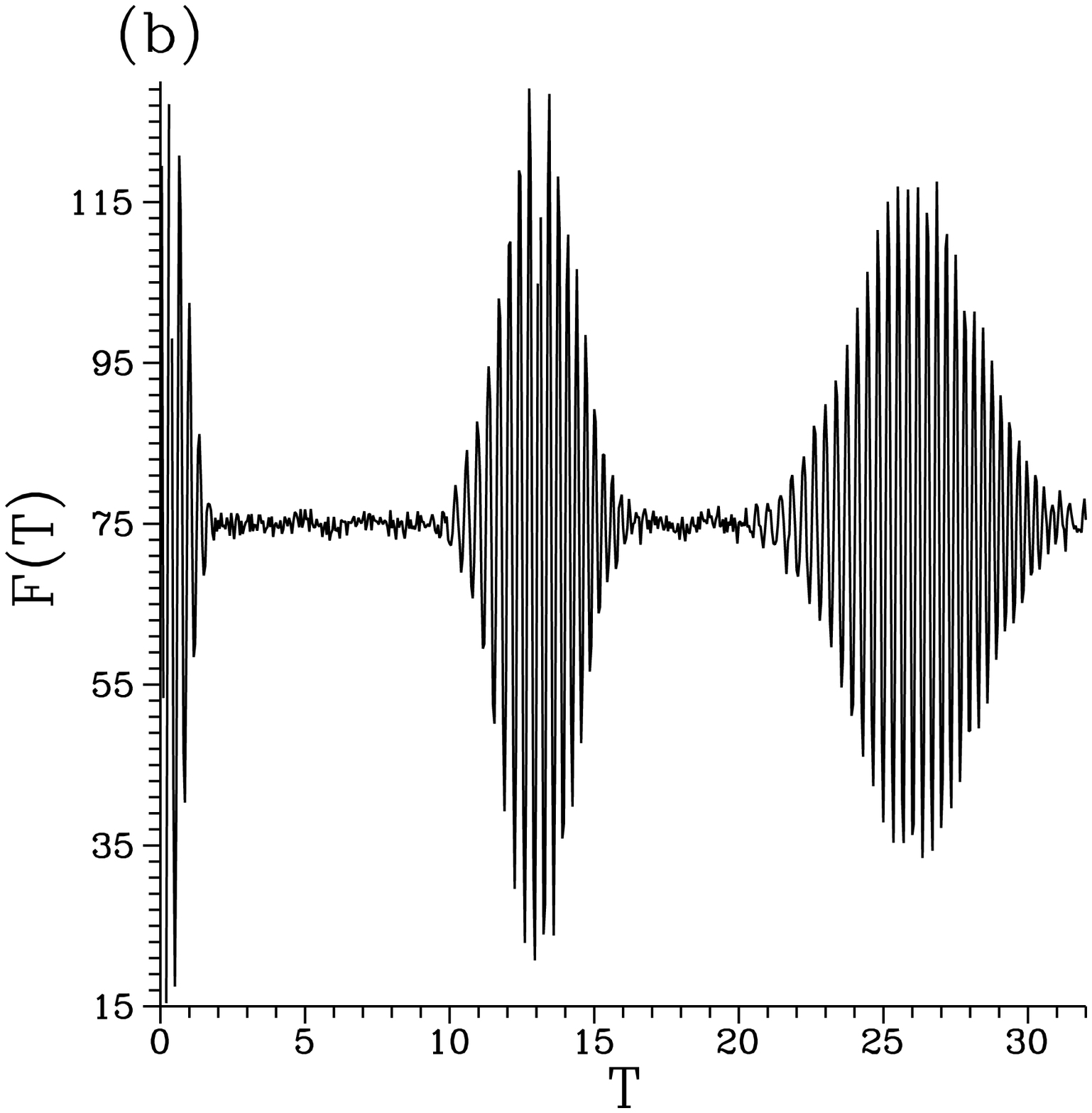}}
 {\includegraphics[width=3cm]{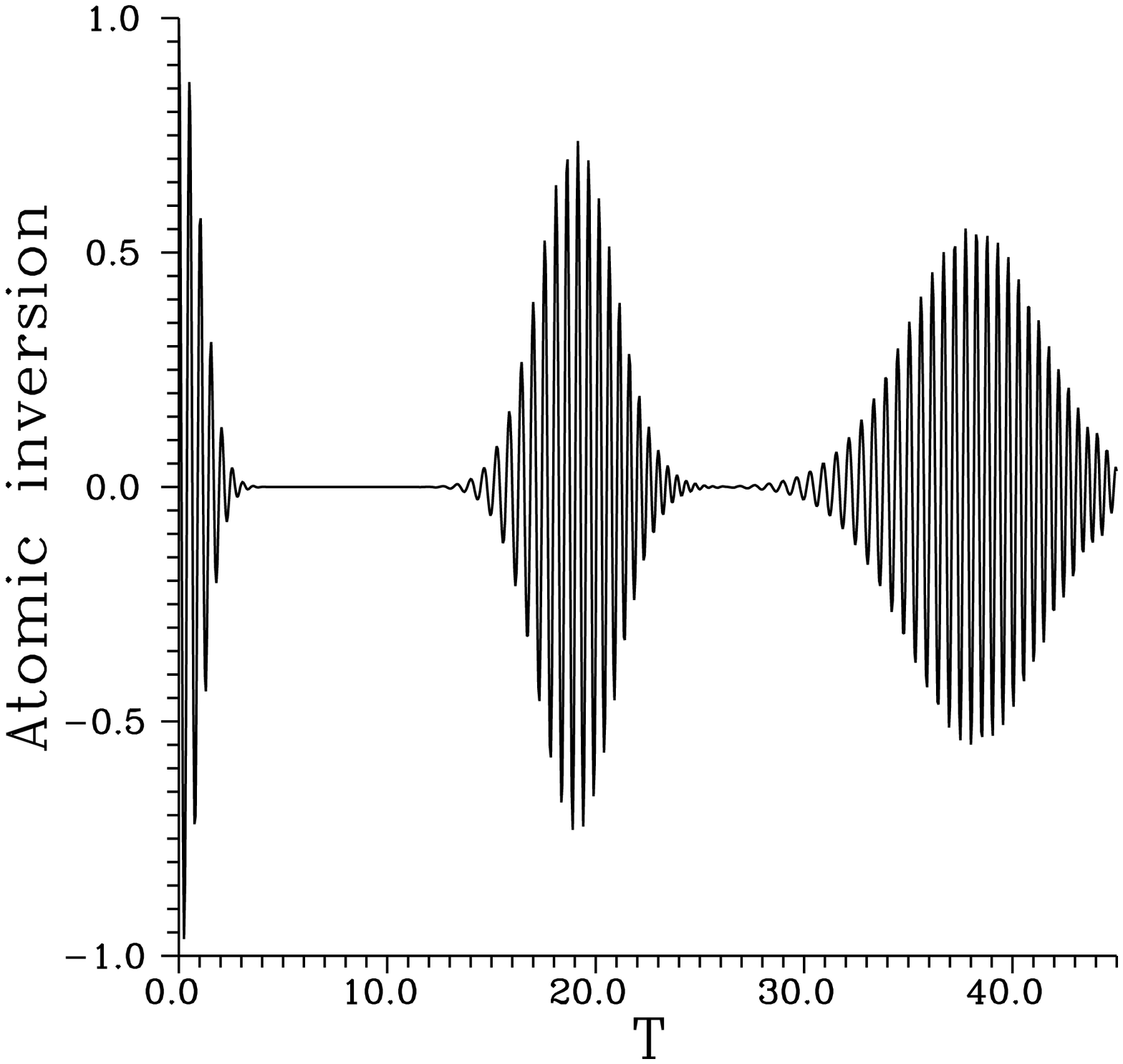}}
 \caption{
The squeezing factor $F$ against $|\alpha|$ (a) and scaled time
$T$ (b) for even coherent states. In  (a) $(k,T)=(1,12.65)$ (solid
curve), $(2,4.5)$ (short-dashed curve) $(3,7.7)$ (long-dashed
curve), $(4,9.6)$ (star-centered curve) and the straight line  is
the squeezing bound. In  (b) $(k,\alpha)=(3,6)$ and
 the inset is the atomic
inversion of the case $(k,\alpha)=(1,6)$.}
\end{figure}
Figure (b) carries information
about the behaviour of the Mandel $Q$ parameter when the optical cavity
field is initially in even coherent states. It is worth reminding that
this state exhibits always super-Poissonian statistics.
However, here through the evolution  we can observe that the
sub-Poissonian statistics are achieved giving their maximum values for $k=1,4$.
In Fig. 8b the curves include smooth oscillations resulting from the superposition
principle.  Further, the sub-Poissonian intervals are increased compared to those
in the Fig. 8a, in particular, for $k=1,2,4$. The two-photon absorption case
(short-dashed curve) almost exhibits steady-state sub-Poissonian
statistics for large values of $|\alpha|$, whereas
 the case $k=3$ exhibits sub-Poissonian statistics only for the large
values of $|\alpha|$.
Finally, figures 8 show that there is no clear relation between the obtained
sub-Poissonian statistics and the values of the absorption parameter $k$.

Now we turn our attention to discuss quadrature squeezing. Actually, it
has been shown that squeezed light can be obtained in the standard JCM
 \cite{meys}.  However, for input squeezed light with
the atom  initially prepared in excited atomic state the interaction destroys
the initial squeezing, which is inherent in the radiation field \cite{mann}.
Here we note that squeezing can be obtained provided that the initial
superimposed optical cavity is squeezed. Moreover, the system
 can switch squeezing from one quadrature to the other. The origin of
 this is in the periodic nature of the system.
For example, even coherent state with real $\alpha$ exhibits squeezing
in $Y$-quadrature provided that $0\leq |\alpha|<2$ \cite{gerr}.
In the present system  squeezing can  be transferred
to $X$-quadrature independent of
the values of $k$ (see Fig. 9a for given values of the
parameters). In this figure the values of the interaction time have
been chosen  numerically in order to maximize the amount of squeezing in
$F$.
We proceed,  from Fig. 9a for each value of $k$,  $F$ starts from almost
minimum-uncertainty state around $|\alpha|=0$ to increase the amount of
 squeezing until it  exhibits
its maximum squeezing, then  starts to decrease the squeezing and eventually
exhibits classical effect (i.e. squeezing  completely disappears).
Also from this figure the maximum (minimum) amount of
squeezing occurs for the case $k=2\quad (1)$ for these chosen values of
the parameters.
   Fig. 9b includes the dynamical behaviour for the
$X$-quadrature for the case $(k,|\alpha|)=(3,6)$. It is obvious that
there is no  nonclassical negative values (, i.e. there is no squeezing).
Nevertheless,
$F(T)$ exhibits collapses and revivals, which (apart from the different scales)
are very  similar to  those of the
$\langle \hat{\sigma}_{z}(T)\rangle$ for the case $k=1$
(compare the behaviour of
the inset in Fig. 9b  with that of $F(T)$).
To be more specific, qualitatively both these curves are equivalent in
such a way that they collapse, remain quiescent, revive, collapse
again and repeatedly undergo a complicated pattern of collapses and revivals for
large interaction time, which is not included in the figure.
Such type of equivalence leads to an interesting result: the collapse-revival
phenomenon in the $\langle \hat{\sigma}_{z}(T)\rangle$ for the standard JCM
can be
detected indirectly via measuring the quadrature squeezing for three-photon
absorption case  through  the homodyne detection.
It is worth mentioning that observation of collapses and revivals were
performed using the one-atom mazer \cite{remp}, which is more sophisticated than the
dynamics of the JCM.

\section{Wigner function}
In this section we investigate the behaviour of the  Wigner ($W$) function for the
system under consideration.
Actually, the $W$ function is informative,  sensitive to the interference
in phase space, and can  give a prediction to  the possible
occurrence of the  nonclassical effects.
Furthermore, this function can  be constructed experimentally
 using the optical homodyne tomography \cite{tom}.

The $W$ function is defined \cite{wolf} as
\begin{equation}
W(x,p,T)=\frac{1}{2\pi}\int_{-\infty}^{\infty} d\zeta\exp(-ip\zeta) \langle x+\frac{\zeta}{2}|
\hat{\rho}_{f}(T)| x-\frac{\zeta}{2}\rangle,\label{11}
\end{equation}
where we have assumed that $\hbar=\omega_{0}=1$.
On substituting the wavefunction obtained from (\ref{8}) into (\ref{11}) and using both the wavefunction of the
Fock state as
\begin{equation}
\langle x|n\rangle=\frac{\exp(-\frac{x^{2}}{2})}{\sqrt{\pi^{\frac{1}{2}}2^{n}n!}}
{\rm H}_{n}(x) \label{ina}
\end{equation}
and the identity \cite{intg}
\begin{equation}
\int_{-\infty}^{\infty}\exp(-\zeta^{2}) {\rm H}_{ m}(\zeta +z_{1})
{\rm H}_{ n}(\zeta +z_{2})d\zeta=2^{n}\sqrt{\pi}m!z_{2}^{n-m}{\rm
L}^{n-m}_{m}(-2z_{1}z_{2}), \quad m\leq n, \label{inb}
\end{equation}
 we arrive at
\begin{eqnarray}
\begin{array}{lr} W(x,p,T)=\frac{\exp(-|\chi|^{2})}{\pi}
\Bigl\{\sum\limits_{ m,m'=0}^{\infty} (-1)^{m'}
\chi^{m-m'}2^{\frac{m-m'}{2}}C_{m}(\alpha ,r ,n,\epsilon)
C^{*}_{m'}(\alpha,r,n,\epsilon)\\
\\
\Bigl[
\cos [T\sqrt{h(m,k)}] \cos [T\sqrt{h(m',k)}]\sqrt{\frac{m'!}
{m!}}{\rm L}^{m-m'}_{m'}(2|\chi|^{2})
 \\
 \\
 +(-1)^{k}\sin [T\sqrt{h(m,k)}] \sin [T\sqrt{h(m',k)}]
\sqrt{\frac{(m'+k)!}
{(m+k)!}}{\rm L}^{m-m'}_{m'+k}(2|\chi|^{2})\Bigr]
\Bigr\},
\end{array}
\label{13}
\end{eqnarray}
where $\chi=x+ip$.

Before  discussing the behaviour of (\ref{13}) it is worth
mentioning that
the $Q$  distribution function of the standard JCM with and without
cavity damping have been investigated in \cite{ris1,{ris2},{ris3},{ris4}}. It has been
shown, for the non-damping case, that the initial shifted Gaussian distribution splits into two
distributions, which counter-rotate on a circle in the complex plane of
the distribution. At the opposite end of the circle the two peaks
collide, they split again, and so forth. This behaviour
is closely connected to the revival-collapse
phenomenon in the $\langle \hat{\sigma}_{z}(T)\rangle$, where
 collapses of the Rabi oscillations occur during splitting of the
distribution and the revivals occur during collision of the peaks of the distribution.
As the system evolves the width of the peaks become broader. When they
 spread over the whole circle no collision occurs anymore and
distinct collapses and revivals are no longer observed.
Also  splitting in the $Q$
distribution for the JCM is explained based on the interpretations of
the density matrix \cite{matsuo}.
Further, it is worth mentioning that the generation of superposition states in the
standard JCM has been investigated using this function.
For the damping
case the counter-rotating peaks spiral into the origin till finally the
stationary distribution is reached. Employing $W$ function
for JCM has stimulated few efforts, in particular, for the damping case
\cite{ris1}.
Here we provide a connection between the evolution of both
$\langle \hat{\sigma}_{z}(T)\rangle$ and  the value  of
the $W$ function at the origin for the two following cases.

Case (i): when the absorption parameter is odd number, i.e.
$k=1,3,5,..$.  From
(\ref{13}) we can write the following expression

\begin{equation}
W(0,T)=\frac{1}{\pi}
\sum\limits_{m=0}^{\infty}(-1)^{m}P(m)
\cos [2T\sqrt{h(m,k)}],  \label{13a}
\end{equation}
where  only the diagonal terms survive.
Careful examination to (\ref{13a}) leads to the following facts:
for the pair-photon states (i.e. even or odd states) (\ref{13a})
can be rewritten as

\begin{equation}
W_{\pm}(0,T)=\pm\frac{1}{\pi}
\langle \hat{\sigma}_{z}(T)\rangle,\label{13b}
\end{equation}
where the positive (negative) sign refers
to even (odd) case. This expression indicates that when
$W(0,T)\simeq 0 \quad (\neq 0)$ the collapse (revival) occurs in
$\langle \hat{\sigma}_{z}(T)\rangle$.
It is obvious that when $W(0,T)=\pm 1/\pi$ the
amplitude  of $\langle \hat{\sigma}_{z}(T)\rangle$ is
maximum. Furthermore, the value  of
$W$ function at the origin exhibits
nonclassical negative values  in the course of revival times,
which means that the nonclassical effects most probable occur during this
time. On the other hand, for the non-pair-photon states
the opposite behaviour occurs.
For instance, for the standard JCM
$W(0,T)\neq 0 \quad (\simeq 0)$ the collapse (revival) occurs in
$\langle \hat{\sigma}_{z}(T)\rangle$.
These ensure that Schr\"{o}dinger-cat state can be generated during the
collapse time, where the establishment of the interference in phase space
is quite obvious.  This fact will be made clear shortly.

Case (ii): when the absorption parameter is even number, i.e.
$k=2,4,...$. In this case the origin of  the $W$
function  is localized in phase space (time independent)
 having its initial value
$W(0,T)=W(0,0)$. This leads to a novel result, which is: the nonclassical
states whose $W$ functions have negative values at the origins
(e.g. odd coherent state)  will be always nonclassical
when they are evolving through the JCM with even absorption parameter.
Of course the chaotic behaviour (and/or overlap of the successive revivals)
in $\langle \hat{\sigma}_{z}(T)\rangle$ can be
recognized in the evolution of the $W$ function at the origin when $|W(0,T_{1})|\simeq |W(0,T_{2})| $
such that $T_{1}\geq T_{2}$.
It is worth referring to the nonlinear version scheme of a
single-atom homodyne detector in which the ionization probability of a
single test atom is given in terms of the Wigner characteristic
function of the field \cite{marin}.
\begin{figure}
{\includegraphics[width=6cm]{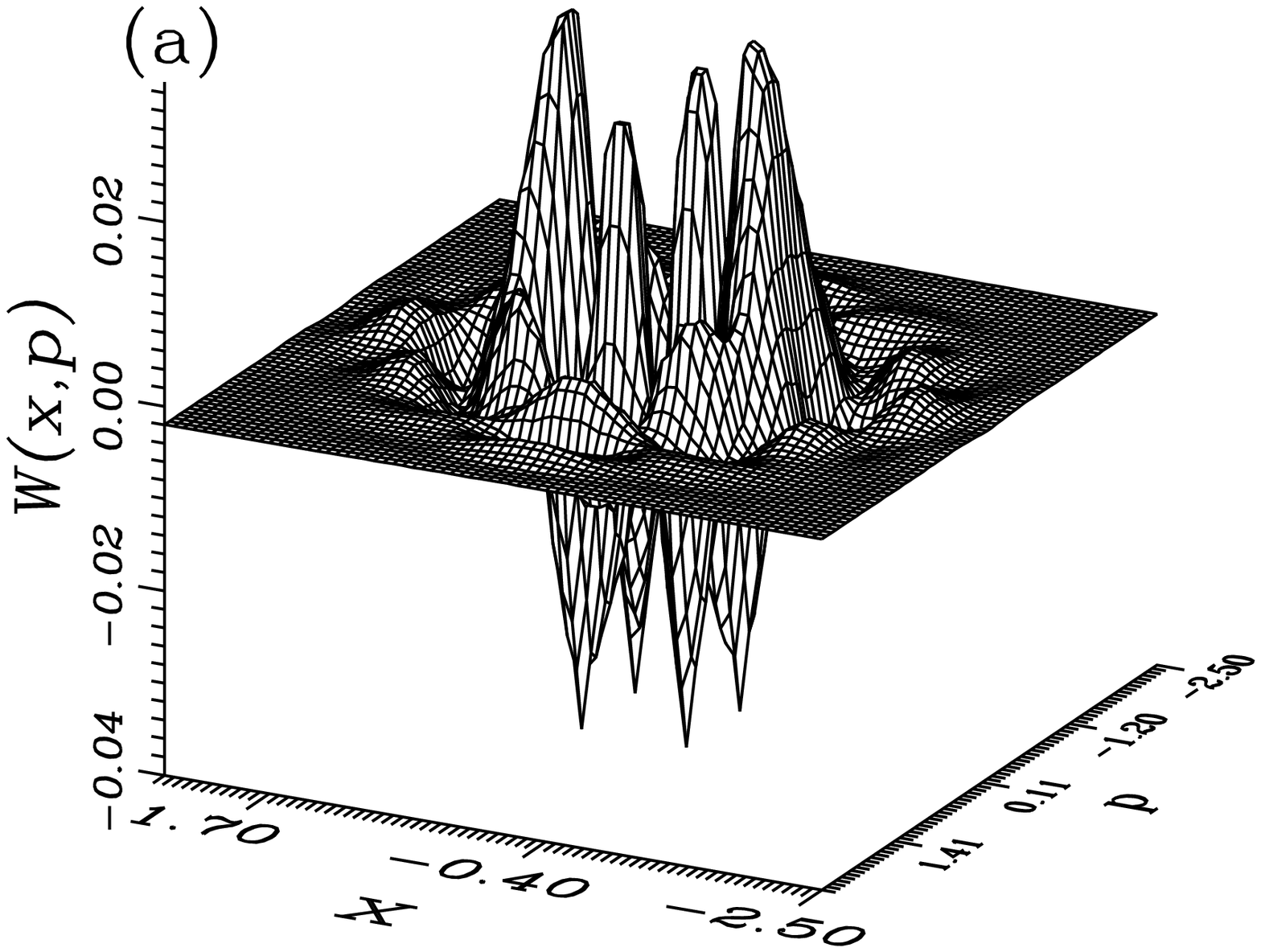}}
{\includegraphics[width=6cm]{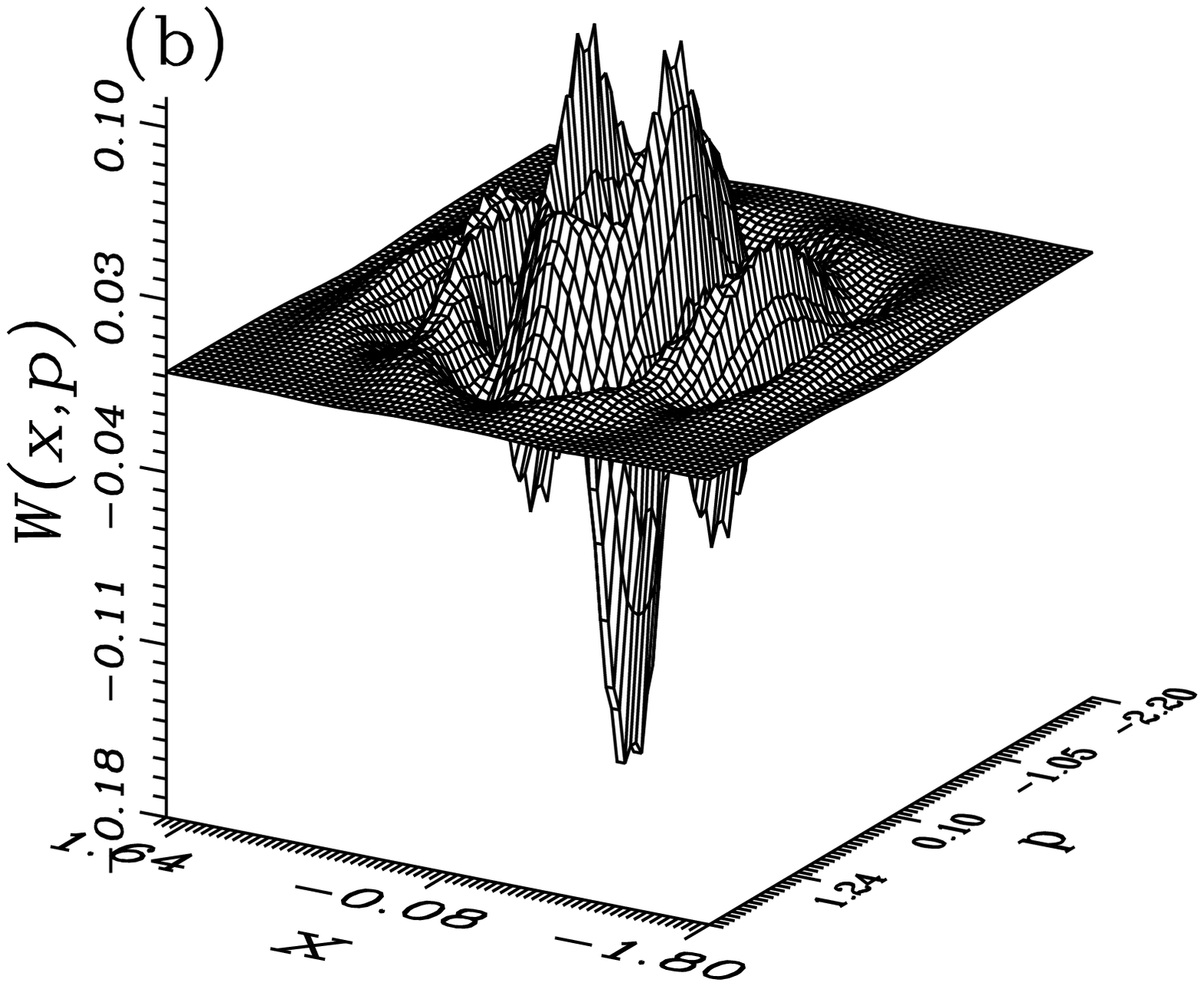}}
\caption{$W$ function of
the even displaced Fock state for $\alpha=3, k=1, n=1$ and for: a)
$T = 4.599998$ (collapse time) and b) $T = 12.60001$ (revival
time).}
\end{figure}
In Figs. 10 a and b we plot $W$ function for even displaced
number state for $T = 4.599998$ (collapse time) and $12.60001$
(revival time), respectively, for the given values of the parameter.
The values of the interaction time are chosen numerically.
Actually, $W$ function of one-photon even coherent state consists of
two inverted holes, with more pronounced negative values, and interference
fringes inbetween. In Figs. 10 according to the evolution of the
dynamical system such shape is destroyed and the $W$ function becomes
more structuralized, which is more pronounced for the collapsed case Fig. 10a.
To be more specific, Fig. 10 a includes multi-peak structure ensuring
the generation of the Schr\"{o}dinger-cat states.
Also for this case we find that $W(0,T)\simeq 0$.
Conversely, Fig. 10 b includes
 more pronounced negative values at the origin of phase space.
 These facts are in a good agreement with the above conclusions.

The second issue we would like to address here is that it has been
shown that cat states can
be generated at one-half of the revival time in the standard JCM
\cite{julio1,{buza},{hlad},{zahe}}.
One of the ways is the investigation of the $Q$ function. Nevertheless,
this function does not include a complete information about the system
 as the $W$ function includes.
 Now we give the form of the $W$ function for such type
of cat states. We do so for the standard JCM, i.e. $k=1$ and
$\epsilon=n=r=0$, with strong intensity
in such a way that the harmonic approximation is applicable. To deal
with  the  second  part of (\ref{13}) some tricks should be
performed, e.g.  replacing $C_{m}$ by $C_{m+1}$ since we are
working in the strong-intensity regime such a transformation is relevant.
 On using (\ref{16}) in addition to the properties of the Laguerre polynomial
 (\ref{13}) reduces to

\begin{eqnarray}
\begin{array}{lr} W(x,p,T)=\frac{1}{2\pi}\exp
[-(x-\varepsilon_{1})^{2}]\Bigl\{
\exp [-(p-\varepsilon_{2})^{2}]+\exp [-(p+\varepsilon_{2})^{2}]\\
\\
-2\exp(-p^{2})\sin(\frac{T\eta_{2}}{2})\sin\left[T(\eta_{1}-\frac{\eta_{2}}{2})+2\alpha
\left(\sqrt{2}x-\alpha\cos(\frac{T\eta_{2}}{2})\right)\sin(\frac{T\eta_{2}}{2})
\right]\Bigr\},
\label{w1}
\end{array}
\end{eqnarray}
where
\begin{equation}
\varepsilon_{1}=\sqrt{2}\alpha \cos(\frac{T\eta_{2}}{2}), \quad
\varepsilon_{2}=\sqrt{2}\alpha \sin(\frac{T\eta_{2}}{2}). \label{w2}
\end{equation}
We should stress that for the best of our knoweldge this is the first time  for which
the exact form of the
$W$ function, i.e. (\ref{w1}), of the generated cat state in the microcavity is
given.
Expression (\ref{w1}) shows that the peaks centers move on the circle
controlled by (\ref{w2}). In fact, the appearance of the
factor $\sqrt{2}$ in the
definition of the circle is related to the definition of $W$ function
(\ref{11}), however, if one uses the transformation $(x,p)\rightarrow
\sqrt{2}(q,y)$ then the relations (\ref{w1}) and (\ref{w2})
would be consistent  with those obtained earlier for the $Q$ function, e.g. \cite{ris4}, provided  that
the prefactor $1/(2\pi)$ is replaced by $1/\pi$. Such a prefactor is
connected with the transformation in the integration  (\ref{11}).
Expression (\ref{w1}) shows  that at the revival
times (i.e. $T=2\pi/\eta_{2}$) the two peaks collide, whereas at
one-half revival time the well-known shape of the cat-state $W$ function
 is obtained, which is two Gaussian bells corresponding
to statistical mixture of individual composite states and
interference fringes inbetween arising from the contribution of the
overlapping between different components of the state.
It is worth referring to \cite{ris4}, in which an analytical closed
form for the $Q$ function is derived by applying the method of steepest
decent and saddle-point integration in the limit of high photon numbers.
The final remark,  the asymptotic form for the $W$ function
related to the two-photon
absorption case can be obtained from  (\ref{w1}) by simply setting
$\eta_{1}=3$ and $\eta_{2}=2$. In this case the $W$ function is intensity
independent and its behaviour is systematic as that of $\langle
\hat{\sigma}_{z}(T)\rangle$.

\section{Conclusions}
In this paper we have discussed the quantum properties for
superposition of squeezed  displaced number states
defined by (\ref{1}) against
the lossless multiphoton Jaynes-Cummings model.
We have assumed that the atom is initially prepared in the excited state.
In addition to the characteristic features of the
 system the investigation gives  generalization to various results in the literatures.
The investigated quantities are the atomic inversion, photon-number
distribution, purity, Mandel $Q$ parameter, quadrature squeezing and $W$
function. The main conclusions are  the followings:
Atomic inversion exhibits revival-collapse pattern as well as echoes,
which are
readout of the photon-number distribution. Furthermore, including squeezing in the
superimposed optical cavity can cause revivals and collapses in the
atomic inversion.
The photon-number distribution shows that various forms of cat states can be generated
via the system.
Also the entanglement and disentaglement between the field and atom
occur periodically
since the system is characterized by the  periodic exchange of energy
between the atom and the cavity mode.
Such behaviour has been confirmed in the purity evolution.
Actually, the behaviour of these quantities
basically depends upon the values of the absorption parameter $k$, which results in the
functional dependence   of the Rabi frequency on the mean photon number.
The sub-Poissonian statistics occur regardless of the values of $k$ provided that
$\alpha$ is finite. Also squeezing in the quadrature squeezing can be
occurred and switched
between the two  quadratures  provided that the
initial superimposed optical cavity is squeezed. Furthermore,
 the evolution of the
quadrature squeezing for the case $k=3$ exhibits  revivals and collapses
similar to those of  the atomic inversion of the standard JCM. Actually,
this is a novel result and its consequence is
 that the revival-collapse phenomenon exhibited for the atomic inversion
 can be measured by homodyne detection via three-photon microcavity
 \cite{elo}.
With respect to $W$ function we have found some novel results, which can
be summarized as follows:
There is  a direct relation  between the evolution of the
values of the $W$ function at the origin in phase space and the evolution of
$\langle \hat{\sigma}_{z}(T)\rangle$
when the absorption parameter $k$ is odd number.
This suggests that  the atomic inversion can be detected
via both the homodyne tomography \cite{tom} and trapped ion \cite{ion1}.
It is worthwhile mentioning that in the trapped ion methods
  a proposal for
measuring the $W$ function at the origin of the phase space for  a
single photon field is given \cite{ion1}  and this would be relevant for the task.
On the other hand, for even absorption  parameter
the origin  of the $W$ function in phase space is localized and thus
 the nonclassical states whose $W$ functions  include negative
values at the origin  will be always nonclassical
 when they are evolving with the JCM.
Furthermore, for the first time, as far as we know, we have given  the explicit form of the
$W$ function of the generated cat state in the microcavity.

\end{document}